\documentclass[twocolumn, nofootinbib, showpacs, superscriptaddress]{revtex4-1} 
\usepackage{amsfonts}
\usepackage{amsmath}
\usepackage{ulem}
\usepackage{dsfont}
\usepackage[pdftex]{graphicx}
\usepackage{epstopdf}
\usepackage{textcomp} 
\usepackage{wasysym} 
\usepackage{graphicx}
\usepackage{bm}
\usepackage{cancel}
\usepackage[varg]{txfonts} 
\usepackage{xcolor}
\usepackage{comment}
\usepackage{hyperref}
\hypersetup{
    colorlinks=true,
    linkcolor=red,
    citecolor=blue,
} 

\specialcomment{supplement}{\begingroup\color{cyan}\footnotesize}{\endgroup}
\specialcomment{formula}{\begingroup\footnotesize}{\endgroup}
\excludecomment{details}

\newcommand{\W}[0]{{\rm{W}}}
\newcommand{\sigp}{{\sigma_{\! \! p}}}
\newcommand{\sigx}{{\sigma_{\! \! x}}}

\newcommand{\Mpc}{{\,\mathrm{Mpc}/h}}

\newcommand{\del}[0]{\partial }
\newcommand{\sH}[0]{{\mathcal{H}}}

\newcommand{\vol}[2]{\hspace{-0.8mm}\mbox{$\text{d}^{\hspace{-0.0mm}#1}$}\hspace{-0.2mm}#2\hspace{0.8mm}\ }

\renewcommand{\d}{\text{d}}
\renewcommand{\v}[1]{\bm{#1} }
\newcommand{\vx}[0]{\bm{x} }

\newcommand{\vp}[0]{\bm{p} }
\newcommand{\vk}[0]{\bm{k} }
\newcommand{\vq}[0]{\bm{q} }
\newcommand{\vnabla}[0]{\bm{\nabla} }

\begin{document}
\title{Coarse-grained cosmological perturbation theory: stirring up the dust model}
\author{Cora Uhlemann} 
\email{cora.uhlemann@physik.lmu.de}
\affiliation{Excellence Cluster Universe, Boltzmannstr. 2, 85748 Garching, Germany} 
\affiliation{Arnold Sommerfeld Center for Theoretical Physics, Ludwig-Maximilians-University, Theresienstr. 37, 80333 Munich, Germany} 
\author{Michael Kopp} 
\email{michael.kopp@physik.lmu.de}
\affiliation{Excellence Cluster Universe, Boltzmannstr. 2, 85748 Garching, Germany} 
\affiliation{Arnold Sommerfeld Center for Theoretical Physics, Ludwig-Maximilians-University, Theresienstr. 37, 80333 Munich, Germany} 
\affiliation{University Observatory, Ludwig-Maximilians-University, Scheinerstr. 1, 81679 Munich, Germany}

\begin{abstract}
We study the effect of coarse-graining the dynamics of a pressureless selfgravitating fluid (coarse-grained dust) in the context of cosmological perturbation theory, both in the Eulerian und Lagrangian framework. We obtain recursion relations for the Eulerian perturbation kernels of the coarse-grained dust model by relating them to those of the standard pressureless fluid model. The effect of the coarse-graining is illustrated by means of power and cross spectra for density and velocity that are computed up to 1-loop order. In particular, the large scale vorticity power spectrum that arises naturally from a mass-weighted velocity is derived from first principles. We find qualitatively good agreement of the magnitude, shape and spectral index of the vorticity power spectrum with recent measurements from N-body simulations and results from the effective field theory of large scale structure. To lay the ground for applications in the context of Lagrangian perturbation theory we finally describe how the kernels obtained in Eulerian space can be mapped to Lagrangian ones.
\end{abstract}

\maketitle
\section{Introduction}
Cold dark matter (CDM) and dark energy comprise 95\% of the energy budget of the Universe and are mainly responsible for its current expansion rate and the observed large scale structure (LSS). The expansion history as well as LSS are thus key to our understanding of the fundamental laws of nature, such that increasingly larger efforts are spent to map them directly or indirectly through galaxy and lensing surveys and through the cosmic microwave background. Therefore accurate analytical models for the formation of LSS are indispensable to constrain cosmological parameters and to search for deviations from $\Lambda$CDM, the standard model of cosmology. In the $\Lambda$CDM model, dark energy is a cosmological constant $\Lambda$ and CDM is a particle species that effectively interacts only gravitationally, or collisionless, and whose initial velocity distribution is cold, or fully described by a smooth   gradient field.

Among analytical methods developed to describe the LSS formation, perturbative schemes based on the popular dust model \cite{P80} play an important part. The dust model describes self-gravitating collisionless cold dark matter (CDM) as a pressureless fluid which fulfills a coupled system of differential equations consisting of continuity, Euler and Poisson equation. These equations can be solved perturbatively -- either in the Eulerian frame \cite{B02} where everything is expanded in terms of density and velocity or in the Lagrangian framework \cite{EB97} where fluid-trajectories or displacements are considered. Those perturbative techniques provide satisfactory results within the linear regime of structure formation and resumming some classes of perturbative corrections \cite{CS06,MP07,M08} can enhance their range of applicability towards mildly nonlinear scales. However, they are condemned to break down eventually in the deeply nonlinear regime due to their inability to dynamically generate higher cumulants like velocity dispersion dynamically. Indeed, the dust model is a truncation of the infinite hierarchy for the cumulants of the phase-space distribution of particles which fulfills the Vlasov (or collisionless Boltzmann) equation. Truncating the hierarchy is only consistent as long as the particle trajectories are well described by a single coherent flow, called single-stream approximation, since as soon as multiple streams become relevant all higher cumulants are sourced dynamically, see \cite{PS09}.

To tackle this shortcoming several semi-analytical methods based on Effective Field Theory (EFT) both in the Eulerian \cite{P11,H12,CHS12,CFGS13,M13,CLP13,M14} and Lagrangian framework \cite{PSZ13,SZ14} have been developed. The strategy of EFT of LSS is to integrate out (or formally solve) the dynamics of the short-wavelength part in order to obtain closed-form equations of motion for the long-wavelength quantities. They describe the large scale physics in terms of an effective fluid that is treated perturbatively and characterized by several parameters arising from small scale physics. These parameters are not calculable within the EFT framework itself but have to be inferred from observations or N-body simulations, at least as long as no full theory describing the small scale physics is at hand. 
All formulations of EFT of LSS have an underlying coarse-graining approach in common but differ in the precise implementation of the cut-off, while some rely on sharp-k filtering, others employ smooth filters like spherical top-hat or Gaussian window functions. The coarse-graining procedure allows to separate long from short scale modes and handle the former perturbatively while regarding the latter as source terms for higher phase space cumulants like velocity dispersion. 

In \cite{D00} a spatially coarse-grained description of a many-body gravitating system for the evolution of LSS has been studied and shown to lead to a fluid-like description which recovers the usual dust model when scales substantially larger than the coarse-graining scale are considered. It was noted that the corresponding hierarchy for the moments can in principle be closed by expressing the microscopic degrees of freedom through the macroscopic density and velocity. This requires the coarse-graining filter to be invertible which excludes sharp-k and top-hat filter but favors the Gaussian window which was considered. In \cite{D00} the Gaussian filter was Taylor-expanded in the filter length $\sigx$ up to leading order, called large-scale expansion. In this expansion the lowest order term was shown to automatically yield the dust model whereas the first order involves a correction proportional to the coarse-graining scale squared $\sigx^2$. It was demonstrated that this term gives rise to a velocity dispersion which enters into the Euler equation and it was argued that it leads to adhesive behavior, see also \cite{BD05}. 
The method described therein was conjectured to allow for successive improvements over the aforementioned dust and adhesion models.

In \cite{P12} a semi-analytical method based on Eulerian perturbation theory was advocated. Therein long and short distance contributions in the Vlasov equation have been separated by introducing a finite spatial coarse-graining scale in order to derive dark matter fluid equations. In this hybrid approach the large scales are treated analytically within perturbation theory while the effect of small scales is included using external source terms to be measured from N-body simulations. As a first step, the perturbative treatment of the sources has been used to illustrate that a velocity dispersion is generated macroscopically by introducing a coarse-graining scale and hence not exclusively caused by shell-crossing effects relevant on small scales. Picking up this idea we will focus on the large scale contribution that can be computed perturbatively. In particular, we will extend previous analyses by taking vorticity, which has been neglected so far, explicitly into account  and by considering a Gaussian filtering which is invertible in contrast to sharp-k and top-hat windows studied so far. A comparison between different filtering schemes used to obtain a truncated Zel'dovich approximation in \cite{M94} revealed that a Gaussian leads to the best agreement with N-body data and considerable improvement over sharp k-truncation as originally suggested in \cite{C93} and top-hat in coordinate space as considered in \cite{P11}.

Our approach is based on the Schr\"odinger method (ScM) as described in \cite{UKH14}, which is able to catch the fully-fledged N-body dynamics and incorporate higher cumulants like velocity dispersion which are relevant for multi-streaming. In the limit $\hbar \rightarrow 0$ the ScM constitutes a full resummation in the filter length $\sigx$ of the coarse-grained hydrodynamics described in \cite{D00}. We will restrict our attention to the mildly nonlinear scales, where shell crossing is not yet dominant. In this regime the limit $\hbar \rightarrow 0$ of the ScM reduces to the coarse-grained dust model which we will study perturbatively in analogy to the dust model. One shortcoming of the dust model is the absence of vorticity and inability to generate it dynamically. While it has been supposed in \cite{PS09} that considering a mass-weighted velocity may account for large scale vorticity, we provide the first consistent implementation of this idea. We compare our result for the vorticity power spectrum to cosmological numerical simulations, see for example \cite{PS09,H14} and the effective field theory of large scale structure \cite{CFGS13}. To lay the foundation for applications of the coarse-grained dust model we describe how  perturbative kernels in the Lagrangian framework can be obtained from those in Eulerian space. 

\paragraph*{Structure}
This paper is organized as follows: In Sec.\,\ref{sec:PS-CDM} we review the phase space description of cold dark matter starting from the Vlasov equation on an expanding background and investigate the hierarchy of moments arising from the Vlasov equation. We then introduce the dust model as well as the coarse-grained dust model and determine the moments of the two different phase space distributions. Sec.\,\ref{sec:pertkernels} is devoted to the derivation of the corresponding Eulerian perturbation kernels to determine the power and cross spectra for the coarse-grained dust model. Furthermore we describe in Sec.\,\ref{sec:Lag} how these kernels can be mapped to Lagrangian space. We conclude in Sec.\,\ref{sec:concl}.

\section{Phase-space description of cold dark matter}
\label{sec:PS-CDM}
\subsection{Vlasov equation}
The dynamics of cold dark matter (CDM) can be conveniently described using a phase space distribution function $f(t,\vx,\vp)$ which contains all relevant information about the system. Imposing phase-space conservation one directly obtains the Vlasov (or collisionless Boltzmann) equation which governs the time evolution of the distribution function. This equation is supplemented by the Poisson equation which encodes gravitational interaction and causes the Vlasov equation to be nonlocal and nonlinear in the phase space distribution $f$. We use comoving coordinates $\vx$ with associated conjugate momentum $\vp=a^2m\ d\vx/dt$, where $a$ is the scale factor satisfying the Friedmann equation of a $\Lambda$CDM or Einstein-de Sitter universe. Then the Vlasov-Poisson system reads
\begin{subequations}
\label{VlasovPoissonEq}
\begin{align}
\label{VlasovEq}
\partial_t f&=   -\frac{\vp}{a^2 m}\cdot\vnabla_{\! \!  x} f + m \vnabla_{\! \!  x} V \cdot\vnabla_{\! \!  p} f \,, \\
&=\left[ \frac{\vp^2}{2a^2m}+m V(\vx)\right] \left( \overleftarrow{\vnabla}_{\! \!  x} \overrightarrow{\vnabla}_{\! \!  p}- \overleftarrow{\vnabla}_{\! \!  p} \overrightarrow{\vnabla}_{\! \!  x}\right) f \,,\\
\Delta V &= \frac{4\pi G\,\rho_0}{a}  \left(\int \vol{3}{p}\!\!f - 1\ \right) \label{PoissonEq}\,,
\end{align}
\end{subequations}
where $\rho_0$ is the (constant) comoving matter background density such that $f$ has a background value or spatial average value $\langle \int \vol{3}{p}\!f\rangle_{\rm vol}=1$. For convenience we will in general suppress the $t$-dependence of the distribution function. 

\subsection{Hierarchy of Moments}
\label{subsec:Hierarchy}
Since the phase space distribution function $f$ depends on seven variables -- three each for position and momentum and one for time -- it is more manageable to consider purely spatial distributions which characterize the system. This can be done by taking moments of the phase space distribution function with respect to momentum. 
\paragraph*{Generating functional for moments and cumulants} The moments $M^{(n)}$ can be conveniently obtained from the generating functional $G[\v{J}]$ by taking functional derivatives. The cumulants $C^{(n)}$ provide an alternative yet equivalent description elucidating the prominent dust-model, the only known consistent truncation of the Vlasov hierarchy. The generating functional, moments and cumulants are given by
\begin{subequations}
\begin{align}
&G[\v{J}] = \int \vol{3}{p} \exp\left[i\vp\cdot\v{J}\right] f(\vx,\vp) \,, \label{genfun}\\
&M^{(n)}_{i_1 \cdots i_n}:=\int \vol{3}{p} p_{i_1} \ldots p_{i_n} f = (-i)^n \left.\frac{\del^n G[\v{J}]}{\del J_{i_1} \ldots \del J_{i_n}} \right|_{\v{J}=0} \,, \label{moments}\\
&C^{(n)}_{i_1 \cdots i_n}:= (-i)^n \left.\frac{\del^n \ln G[\v{J}]}{\del J_{i_1} \ldots \del J_{i_n}} \right|_{\v{J}=0} \label{cumulants}\,.
\end{align}
\end{subequations}
\paragraph*{Evolution equations} The dynamics of the moments $M^{(n)}$ are encoded in the Vlasov equation \eqref{VlasovEq} and can be extracted easily
\begin{align}
\label{VlasovHierarchy}
\partial_t M^{(n)}_{i_1 \cdots i_n} &=   - \frac{1}{a^2 m} \nabla_j M^{(n+1)}_{i_1 \cdots i_n j}  - m \nabla_{(i_1} V \cdot M^{(n-1)}_{i_2 \cdots i_n)} \,,
\end{align}
where indices enclosed in round brackets imply symmetrization according to $a_{(i}b_{j)}=a_ib_j+a_jb_i$. Unfortunately, the time-evolution of the $n$-th moment depends in turn on the $n+1$-th moment thereby constituting an infinite coupled hierarchy. The underlying hierarchy becomes much more transparent when expressed in terms of cumulants $C^{(n)}$
\begin{align}
\notag \del_t C^{(n)} _{i_1\cdots i_n} &= -\frac{1}{a^2m} \Bigg\{ \nabla_j C^{(n+1)}_{i_1\cdots i_n j} +\sum_{S\in \mathcal P(\{i_1,\cdots,i_n\})} C^{(n+1-|S|)}_{l\notin S,j} \cdot \nabla_j C^{(|S|)}_{k\in S} \Bigg\}\\
&\quad - \delta_{n1} \cdot m \nabla_{i_1}V \,,
\label{cumeq}
\end{align}
where $S$ runs through the power set $\mathcal P$ of indices $\{i_1,\cdots,i_n\}$ and the Kronecker $\delta_{n1}$ in the last term ensures that the potential contributes only to the equation for the first cumulant describing velocity.

\paragraph*{Strategies for closing the hierarchy}
Describing the physical system analytically in terms of a small number of degrees of freedom demands either (i) truncating the hierarchy by ignoring higher cumulants or setting them to zero, or (ii) resorting to a special ansatz for the distribution function. The prominent dust model is an example which combines (i) and (ii) by providing the only known consistent truncation of Eq.\,\eqref{cumeq}. At second order one can set $C^{(n\geq 2)} \equiv 0 $ in a consistent manner since each summand in the evolution equation of $C^{(2)}$, Eq.\,\eqref{cumeq}, contains a factor of $C^{(n\geq 2)}$. At higher order it is evident from Eq.\,\eqref{cumeq} that a truncation is a priori not possible. Numerical studies indicate that as soon as velocity dispersion encoded in $C^{(2)}$ becomes relevant, all higher cumulants are sourced dynamically, see \cite{PS09}. Postulating an ansatz for the velocity dispersion -- for example an imperfect fluid -- or introducing an artificial adhesion-term in the evolution equation for the velocity, see \cite{SZ89}, corresponds to (i). In this case it is difficult to assess whether one is actually still modeling collisionless matter described by the Vlasov hierarchy of cumulants. In \cite{UKH14} we established an approach relying on (ii), namely the Schr\"odinger method (ScM) which provides an ansatz $\bar f_\W$ for the distribution function. The ScM incorporates higher cumulants that approximately solve the Vlasov hierarchy in a controlled manner and allows to compute them analytically. In the following we will consider the coarse-grained dust model $\bar f_\d$ which can also be obtained from this coarse-grained Wigner ansatz $\bar f_\W$ when sending $\hbar \rightarrow 0$. 

\subsection{Dust model}
Within the dust model CDM is described as a pressureless fluid with density $n(\vx)$ and an irrotational fluid velocity $\vnabla \phi(\vx)$. The velocity remains single-valued at each point meaning that particle trajectories are not allowed to cross and velocity dispersion cannot arise. This regime is usually referred to as `single-stream' indicating that this model breaks down as soon as `shell-crossings' occur and multiple streams become relevant. The corresponding distribution function is
\begin{align}
\label{fdust}
f_\d (\vx,\vp)&= n(\vx) \delta_{\mathrm D}\Big(\vp-\vnabla \phi(\vx) \Big) \;.
\end{align}

\paragraph*{Moments and cumulants}
The generating functional for the dust model where $f_\d$ was inserted in \eqref{genfun} yields
\begin{align}
&G_\d[\v{J}] =n \exp\left[i\vnabla \phi\cdot\v{J}\right] \,.
\end{align}
The moments $M^{(n)}$ and cumulants $C^{(n)}$ are then given by
\begin{subequations}
\begin{align}
M^{(0)} &= n\,, \qquad {M}^{(1)}_i=n \phi_{,i}  \,, \quad M^{(n \geq 2)}_{i_1\cdots i_n} = n \phi_{,i_1} \cdots \phi_{,i_n} \,, \label{dustmoments}\\
C^{(0)} &= \ln n\,, \quad \ {C}^{(1)}_i=\phi_{,i}  \,, \quad \quad \  {C}^{(n \geq 2)}_{i_1\cdots i_n} = 0 \label{dustcumulants}\,,
\end{align}
\end{subequations}
where the shorthand notation $\phi_{,i}:=\nabla_i\phi$ has been used.
All cumulants of order two and higher vanish identically since the exponent of the generating functional is linear in $\v{J}$. This simply shows that the dust model does not include multi-streaming effects like velocity dispersion, which is encoded in the second cumulant. 

\paragraph*{Evolution equations} 
Due to the absence of higher cumulants in the dust model, the Vlasov equation is equivalent to its first two equations of the hierarchy of moments. The hydrodynamical system consisting of the continuity and Bernoulli equation for a perfect pressureless fluid with density $n$ and velocity potential $\phi/m$ is supplemented by the Poisson equation
\begin{subequations} \label{dustEqs}
\begin{align} 
\del_t n &= -\frac{1}{m a^2}\vnabla\cdot \left(n \vnabla \phi\right)\label{conti} \,,\\
\del_t \phi &= -\frac{1}{2 a^2 m} \left(\vnabla\phi\right)^2 -mV \label{euler}\,,\\
\Delta V&=\frac{4\pi G\,\rho_0}{a}\Big(n -  1 \Big) \label{Poissdust}  \,.
\end{align}
\end{subequations}
If $n$ and $\phi$ fulfill these equations then all evolution equations \eqref{VlasovHierarchy} of the higher moments are automatically satisfied. By defining an irrotational velocity $\v{u}=\vnabla\phi/m$ one can rewrite \eqref{conti} and \eqref{euler} in the following equivalent form supplemented by the curl-free constraint
\begin{subequations}\label{Fluideq}
\begin{align}
\del_t n &= -\frac{1}{a^2}\vnabla \cdot(n\v{u}) \,, \label{conti2}\\
\del_t \v{u} &= -\frac{1}{a^2}(\v{u}\cdot\vnabla)\v{u} -\vnabla V \,,\label{euler2}\\
\vnabla \times \v{u} &= 0 \,. \label{dustconstr}
\end{align}
\end{subequations}

\subsection{Coarse-grained dust model}

The distribution function of the coarse-grained dust model is defined as a smoothing of the dust probability distribution \eqref{fdust} with a Gaussian filter of width $\sigx$ and $\sigp$ in $\vx$ and $\vp$ space, respectively. For convenience we will adopt the shorthand operator representation of the smoothing which can be easily obtained by switching to Fourier space 
\begin{align}
\label{fcgdust}
\notag \bar f_\d &= \int \frac{\vol{3}{x'}\vol{3}{p'}}{(2 \pi\sigx\sigp)^3 } \exp\left[-\frac{(\vx-\vx')^2}{2\sigx^2}-\frac{(\vp-\vp')^2}{2\sigp^2} \right] f_\d(\vx',\vp')\,,\\
\bar f_\d  &=\exp\left(\tfrac{1}{2}\sigx^2\Delta_x+\tfrac{1}{2}\sigp^2\Delta_p\right) f_\d \,.
\end{align}
As mentioned in the introduction the coarse-grained dust model is a special case of the ScM presented in \cite{UKH14} which can be obtained from the coarse-grained Wigner function $\bar f_W$ in the limit $\hbar \rightarrow 0$ as long as no shell-crossing has occured yet
\begin{align}
\bar f_\W(\vx,\vp) &\stackrel{\hbar \rightarrow 0}{=} \bar f_\d (\vx,\vp) \,. 
\end{align}

If $x_{\rm typ}$ and $p_{\rm typ}$ are the (minimal) scales of interest we have to ensure that
\begin{align}
\sigx \ll x_{\text{typ}} \quad \mathrm{and} \quad \sigp \ll p_{\text{typ}} \,.
\end{align}

\paragraph*{Moments and cumulants}
The generating functional for the coarse-grained dust model is given by
\begin{align}
\label{genfuncgW}
\bar G_\d[\v{J}] &= \exp\left(\tfrac{1}{2}\sigx^2\Delta-\tfrac{1}{2} \sigp^2\v{J}^2\right) G_\d[\v{J}]  \,.
\end{align}
From this expression the calculation for the moments $\bar M^{(n)}$ is straightforward and shows that the first two are given by a spatial coarse-graining of the dust moments \eqref{dustmoments}
\begin{subequations} \label{cgdustmoment}
\begin{align}
\bar M^{(0)}&= \exp\left(\tfrac{1}{2} \sigx^2\Delta\right)  M^{(0)} =: \bar n  \label{cgdustmoment0} \quad , \quad \bar C^{(0)}=\ln \bar n \\
\bar M^{(1)}_i &=\exp\left(\tfrac{1}{2} \sigx^2\Delta\right)  M^{(1)}_i =: m\bar n\bar u_i  \ ,\  \bar C^{(1)}_i = m\bar u_i \label{cgdustmoment1} \,.
\end{align}
The macroscopic velocity $\bar{\v{u}}$ is the mass-weighted dust velocity which is obtained by smoothing the momentum field $n \v{u}$ and then dividing by the smoothed density field $\bar{n}$. This is precisely the definition commonly used in the EFT of LSS, compare \cite{M13, B10}. From a physical point of view $\bar{\v{u}}$ describes the center-of-mass velocity of the collection of particles inside a coarsening cell of diameter $\sigx$ around $\v{x}$.\\
Note that higher moments $\bar M^{(n\geq 2)}$ are not simply given by the coarse-graining of $ M^{(n\geq 2)}$ but receive an extra $\sigp^2$-term 
\begin{align}
\bar M^{(2)}_{ij} &=    \exp\left(\tfrac{1}{2}\sigx^2 \Delta\right) \left\{ M^{(2)}_{ij} +\sigp^2 M^{(0)} \delta_{ij} \right\}  \label{cgdustmoment2}\,,\\
\bar M^{(3)}_{ijk}&= \exp\left(\tfrac{1}{2}\sigx^2 \Delta\right) \left\{  M^{(3)}_{ijk}+ \stackrel{+\text{cyc. perm.}}{\sigp^2 M^{(1)}_i \delta_{jk} } \right\}   \,. \label{cgdustmoment3}
\end{align}
 The corresponding cumulants can be calculated from the previous results using
\begin{align}
\bar C^{(2)}_{ij} 
&= \sigp^2 \delta_{ij} + \frac{\overline{n\phi_{,i}\phi_{,j}}}{\bar n} - \frac{\overline{n\phi_{,i}}\ \overline{n\phi_{,j}}}{\bar n^2} \label{barC2} \,,\\
\bar C^{(3)}_{ijk} 
&= \frac{\bar M^{(3)}_{ijk}}{\bar M^{(0)}} - \stackrel{+ \text{cyc. perm.}}{\bar C^{(2)}_{ij}\bar C^{(1)}_k}- \bar C^{(1)}_i\bar C^{(1)}_j\bar C^{(1)}_k \,,
\end{align}
\end{subequations}
with the shorthand notation $\overline{n\phi_{,i}\phi_{,j}}:=\exp\left(\tfrac{1}{2}\sigx^2\Delta\right)\left\{ n\phi_{,i}\phi_{,j} \right\}$ and $\overline{n\phi_{,i}}:=\exp\left(\tfrac{1}{2}\sigx^2\Delta\right)\left\{ n\phi_{,i}\right\}$.
We can observe that all higher moments are determined self-consistently from the lowest two, which are dynamical and represent the coarse-grained density $\bar n$ and mass-weighted velocity $\bar{\v{u}}$, respectively. 

\paragraph*{Evolution equations}
The dust equations \eqref{dustEqs} can be employed to obtain evolution equations for the first two moments $\bar n = \bar M^{(0)}$ and $\bar u_i=\bar M^{(1)}_i/(m\bar n) $ corresponding to coarse-grained density and mass-weighted velocity, respectively
\begin{subequations}
\label{Fluidcg}
\begin{align}
\del_t \bar n &=   -\frac{1}{a^2}\v{\nabla}\cdot(\bar n \bar{\v{u}}) \,, \label{Conticg}\\
\del_t (\bar n \bar u_i) &= -\exp\left(\tfrac{1}{2} \sigx^2\Delta\right) \Bigg\{\frac{1}{a^2 m^2}\nabla_j \left[n\phi_{,i}\phi_{,j} \right] +  n\ \nabla_i V  \Bigg\}\,. \label{Eulercg}
%
\end{align}
Note that $\sigp$ drops out and that one would obtain exactly the same evolution equations when taking moments of the coarse-grained Vlasov equation, see  Eq.(12) in \cite{UKH14} and inserting the moments \eqref{cgdustmoment} for the coarse-grained dust ansatz $\bar f_{\rm d}$, Eq.\,\eqref{fcgdust}. This is because $\bar f_{\rm d}$ fullfils the coarse-grained Vlasov equation before shell crossing. A specific feature of the Gaussian filter we employed here is that it can be inverted such that there exists a closed-form analogue of Eq.\,\eqref{Eulercg} for the macroscopic quantities $\bar n$ and $ \v{\bar u}$. This equation should be valid even after shell crossing. For details we refer the interested reader to \cite{UKH14}.
Since the macroscopic velocity $\v{\bar u}$ is obtained from the dust velocity $\v{u}=\vnabla \phi/m$ by mass-weighting, these equations are supplemented by the constraint 
 \begin{equation}
 m\, \bar n\, \bar{\v{u}} = \exp\left(\tfrac{1}{2}\sigx^2\Delta\right) \left( n  \vnabla \phi \right)\,. \label{Husimiconstr}
  \end{equation}
\end{subequations}
which is the analogue of the curl-free constraint $\v{u}=\vnabla \phi/m$ Eq.\,\eqref{dustconstr} and enforces a very particular non-zero vorticity for $\bar{\v{u}}$.
\label{3rdmoment}
For practical applications, instead of solving the coarse-grained fluid equations \eqref{Fluidcg} for $\bar n$ and $\bar {\v{u}}$ one can simply solve \eqref{dustEqs} for $n$ and $\phi$ and construct the cumulants of interest according to \eqref{cgdustmoment}                                                                                                                                                                                                                                                                                                                                                                                                         . Note that Eqs.\,\eqref{Fluidcg} are naturally written in terms of the macroscopic momentum $\bar{\v{j}} \equiv \bar n \bar{\v{u}}$.

\section{Eulerian Perturbation theory}
\label{sec:pertkernels}
In order to obtain a solution to Eq.\,\eqref{Fluidcg} for the coarse-grained density contrast $\bar \delta=\bar n-1$ and the mass-weighted velocity $\v{\bar v}$, we can solve the microscopic system \eqref{dustEqs} for $\delta=n-1$ and $\v{v}=\vnabla \theta/\Delta$ where $\theta=\Delta \phi/am$ and then simply coarse-grain the result according to
\begin{subequations}
\label{macrofrommicro}
\begin{align}
\bar \delta &= \exp \left(\tfrac{1}{2} \sigx^2\Delta\right) \delta\,, \label{cgdelta} \\
(1+\bar \delta)\v{\bar v} &= \exp \left(\tfrac{1}{2} \sigx^2\Delta\right) \left[(1+\delta)\v{v}\right]\,. \label{cgv}
\end{align}
\end{subequations}
As mentioned before, this procedure is possible as long as the solution space of the ``microscopic'' functions $\delta, \theta$ allows to invert the Gaussian smoothing operation; this it what justified the use of Eq.\,\eqref{Eulercg} instead of the closed equation (45b) for the macroscopic quantities given in \cite{UKH14} in the limit $\hbar\rightarrow 0$. However, at shell-crossing, where $\delta$ diverges at point-, line-  or sheetlike structures a deconvolution is impossible. Therefore considering the coarse-grained dust case does not allow us to genuinely go beyond shell crossing. However, microscopic vorticity and velocity dispersion contribute to the true  macroscopic vorticity $\bar{\v{w}}$ and velocity dispersion  $\bar{C}^{(2)}_{ij}$. Those microscopic contributions simply add to the corresponding quantities of the coarse-grained dust model that arise without any microscopic origin \cite{P12}. Therefore one might hope that  coarse-grained dust captures some aspects of the true macroscopic $\bar{\v{w}}$ and  $\bar{C}^{(2)}_{ij}$.

\subsection{Eulerian kernels for density and velocity}
We write the solution of the coarse-grained dust model \eqref{macrofrommicro} as a perturbative series in Fourier space and expand in terms of the scale factor $a(\tau)$ for the fastest growing mode where conformal time $\tau$ is given by $dt=a(\tau)d\tau$
\begin{subequations}
\label{pertexpcg}
\begin{align}
\bar\delta(\tau,\v{k})&=\sum_{n=1}^\infty a^n(\tau) \bar \delta_n(\v{k}) \,,\label{pertexpdeltacg}\\
 \v{\bar v}(\tau,\v{k})&=\sH(\tau) \sum_{n=1}^\infty a^{n}(\tau) \v{\bar v}_n(\v{k}) \,. \label{pertexpvcg}
\end{align}
\end{subequations}
To obtain formal solutions we proceed along the lines of standard Eulerian perturbation theory described in \cite{JB94}. The general solution may be written in terms of Fourier kernels
\begin{subequations} 
\label{deltavn}
\begin{align}
\notag \bar \delta_n(\v{k})&=\int \frac{\d^3p_1\ldots \d^3p_n}{(2\pi)^{3(n-1)}}\delta_{\rm D}(\vk-\v{p}_{1\cdots n}) \bar F_n(\v{p}_1,\ldots,\v{p}_n) \times \\
&\qquad\qquad \qquad\qquad \times  \delta_1(\v{p}_1)\cdots\delta_1(\v{p}_n)\label{deltan}\,,\\
\notag\v{\bar v}_n(\v{k})=&i\int \frac{\vol{3}{p_1}\cdots\ \vol{3}{p_n}}{(2\pi)^{3(n-1)}}\,\delta_{\rm D}(\v{k}-\v{p}_{1\ldots n}\,)\bar{\v{V}}_n(\v{p}_1,\ldots,\v{p}_n)\times \\
&\qquad\qquad \qquad\qquad \times\delta_1(\v{p}_1)\cdots\delta_1(\v{p}_n)\,. \label{vn}
\end{align}
It is convenient to decompose the velocity $\v{\bar v}$ into velocity divergence $\bar\theta:= \v{\nabla}\cdot \bar{\v{v}}$ and vorticity $\bar{\v{w}}:=\v{\nabla}\times \bar{\v{v}}$ for which we also define Fourier kernels according to
\begin{align}
\notag \bar \theta_n(\v{k}\,)=&-\int \frac{\vol{3}{p_1}\cdots\ \vol{3}{p_n}}{(2\pi)^{3(n-1)}}\,\delta_{\rm D}(\v{k}-\v{p}_{1\ldots n}\,) \bar{G}_n(\v{p}_1,\ldots,\v{p}_n) \times \\
&\qquad\qquad \qquad\qquad \times  \delta_1(\v{p}_1)\cdots\delta_1(\v{p}_n)\label{thetan}\,,\\
\notag\v{\bar w}_n(\v{k}\,)=&-\int \frac{\vol{3}{p_1}\cdots\ \vol{3}{p_n}}{(2\pi)^{3(n-1)}}\,\delta_{\rm D}(\v{k}-\v{p}_{1\ldots n}\,) \bar{\v{W}}_n(\v{p}_1,\ldots,\v{p}_n)\times \\
&\qquad\qquad \qquad\qquad \times\delta_1(\v{p}_1)\cdots\delta_1(\v{p}_n)\,. \label{wn}
\end{align}
\end{subequations}
The corresponding kernels are related to those of velocity $\v{\bar V}_n$ via $\bar G_n= \v{k}\cdot \bar{\v{V}}_n $ and $\bar{\v{W}_n}= \v{k}\times \bar{\v{V}}_n $. In the  dust model the vorticity encoded in $\v{W}_n$ vanishes identically. The recursion relations for the kernels of the microscopic density $F_n$ and velocity divergence $G_n$ are given in Eqs.\,(10) in \cite{JB94}.

Since the macroscopic density contrast $\bar \delta$ is trivially related to the microscopic $\delta$, see Eq.\,\eqref{cgdelta} we have that
\begin{subequations}
\label{cgrecursrel}
\begin{align}
\bar F_n &=\exp\left(-\tfrac{1}{2}\sigx^2 k^2\right) F_n \label{barFn}\,.
\end{align}
where  $F_n$ are the standard SPT kernels for the dust model. Therefore, in Eulerian perturbation theory, the matter power spectrum for the coarse-grained dust model is simply given by the coarse-graining of the dust power spectrum, see Eq.\,\eqref{PSdeltacg}.

In order to determine the coarse-grained velocity field we have to expand Eq.\,\eqref{cgv} perturbatively which gives
\begin{align} \label{cgvpert}
\v{\bar v}_n&=\exp\left(\tfrac{1}{2}\sigx^2\Delta\right) \v{v}_n+\sum_{m=1}^{n-1} \left\{\exp\left(\tfrac{1}{2}\sigx^2\Delta\right)\left( \delta_m  \v{v}_{n-m}\right) - \bar \delta_m  \v{\bar v}_{n-m}\right\}\,,
\end{align}
where $\v{v}_n = \vnabla \theta_n/\Delta$ is the microscopic velocity. Note that the curly bracket in \eqref{cgvpert} basically calculates the difference between the average of a product and the product of averages (this statement is exact at second order). It is precisely this deviation that sources the vorticity $\v{\bar w}_n=\vnabla \times \v{\bar v}_n$ which becomes relevant at second order. In the limit $\sigx\rightarrow0$, this contribution vanishes identically at all orders such that the velocity remains a gradient field thereby recovering the standard SPT kernels from \cite{JB94} for $\sigx\rightarrow0$. The kernels $\bar{\v{V}}_n$ for the velocity $\v{\bar v}$ can be read off from \eqref{cgvpert} 
\begin{align} \label{Vkernel}
\bar{\v{V}}_n(\v{p}_1,\ldots,\v{p}_n) &= \frac{\v{k}}{k^2} \exp\left(-\tfrac{1}{2}\sigx^2 k^2\right) G_n \\
\notag &+\sum_{m=1}^{n-1} \Bigg\{ \exp\left(-\tfrac{1}{2}\sigx^2 k^2\right)F_m \frac{\v{k}_2}{k_2^2} G_{n-m} - \bar F_m \bar{\v{V}}_{n-m}  \Bigg\} \,.
\end{align}
\end{subequations}

Note that the kernel $\bar G_n$  of $\bar \theta=\vnabla \cdot \v{\bar v}$ is not simply given by the coarse-graining of the kernel $G_n$ of $\theta= \Delta \phi / am$ since the velocity is mass-weighted according to \eqref{cgv} . However, at first order we recover a curl-free velocity $\v{\bar v}_1= \v{\nabla} \bar \theta_1/\Delta$ with $\bar\theta_1(\vk) =  - \bar\delta_1(\vk)  $.
\subsection{Power and cross spectra}
In order to check whether our new kernels give sound results in perturbation theory, we calculate here some power and cross spectra up to one-loop order. The full expressions which are displayed in App.\,\ref{AppPower} are convergent and reduce to the known results in the limit where $\sigx \rightarrow 0$. The most interesting result is the power spectrum for the vorticity $\v{w}$ given in Eq.\,\eqref{PSvort}, which vanishes identically in the standard dust model.
The power spectra $P(k)$ corresponding to density $\delta$, velocity divergence $\theta$ and vorticity $\v{w}$ are defined according to
\begin{subequations}
\begin{align}
\langle \delta(\vk)\delta(\vk')\rangle &= (2\pi)^3\delta_D(\vk+\vk')P_{\delta\delta}(k) \,, \\
\langle \theta(\vk)\theta(\vk')\rangle &= (2\pi)^3\delta_D(\vk+\vk')P_{\theta\theta}(k) \,, \\
\langle \v{w}(\vk)\cdot\v{w}(\vk')\rangle &= (2\pi)^3\delta_D(\vk+\vk')P_{\v{w}\v{w}}(k) \,.
\end{align}
Furthermore we have the cross spectrum between density $\delta$ and velocity divergence $\theta$
\begin{align}
\langle \delta(\vk)\theta(\vk')\rangle &= (2\pi)^3\delta_D(\vk+\vk')P_{\delta\theta}(k)  \,.
\end{align}
\end{subequations}
The velocity power spectrum is defined accordingly 
\begin{subequations}
\begin{align}
\langle \v{v}(\vk)\cdot\v{v}(\vk')\rangle &= (2\pi)^3\delta_D(\vk+\vk')P_{\v{v}\v{v}}(k)  \,.
\end{align}
Since $\v{v} = (\vnabla \theta- \vnabla\times \v{w})/\Delta$ it can be easily obtained from the divergence $\theta= \vnabla\cdot \v{v}$ and vorticity $\v{w}=\vnabla\times\v{v}$ power spectra
\begin{align}
k^2 P_{\v{v}\v{v}}(k) &= P_{\theta\theta}(k)+ P_{\v{w}\v{w}}(k) \,.
\end{align}
\end{subequations}

\begin{figure}[t!]
\includegraphics[width=0.47\textwidth]{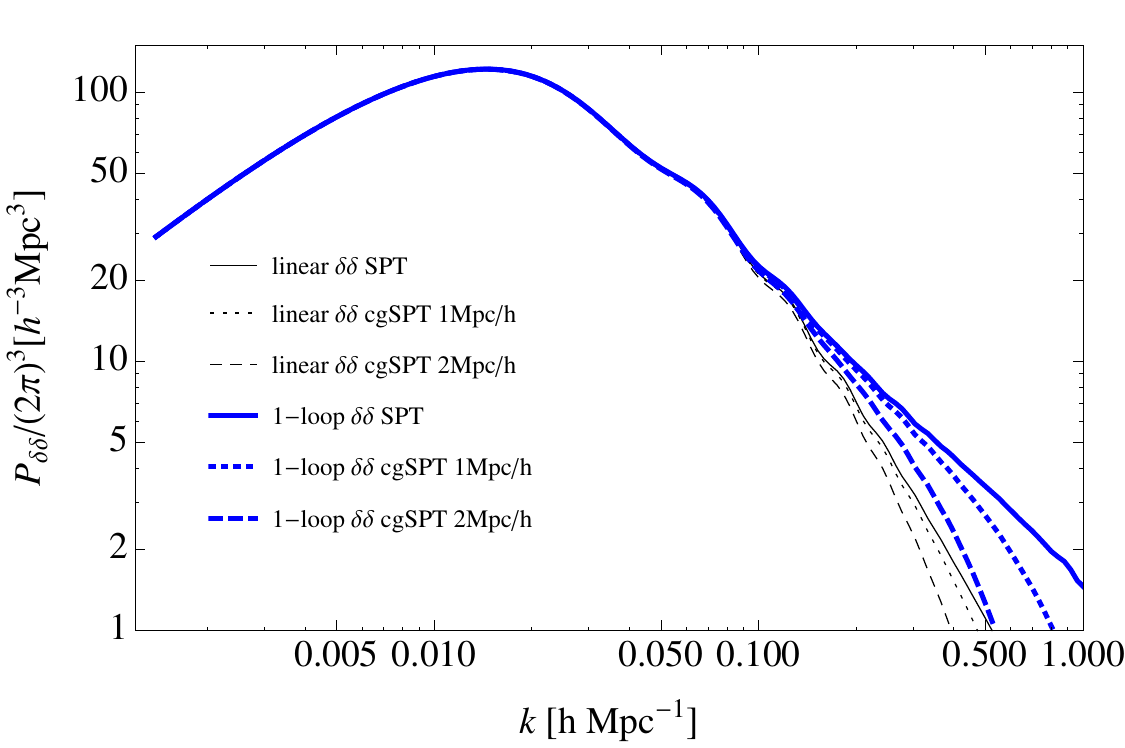}
\caption{Comparison between matter density power spectrum for SPT and cgSPT in 1st (lin) and 2nd (1-loop) order perturbation theory.}
\label{ddspectrum}
\end{figure}

In the following we will derive the power and cross spectra up to one-loop order for the coarse-grained dust model (cgSPT), and compare it to both standard SPT as well as standard SPT with a different coarse-graining procedure (SPTcg) where only the linear input power spectrum $P_L$ is smoothed. This is done merely to illustrate the effect of the coarse-graining on the perturbation kernels rather than to suggest an improvement of SPT. SPT is known to fail to converge as a perturbative series, see \cite{BGK13} and is less accurate in predicting the nonlinear density field than Lagrangian methods \cite{Ta14}. To lay the ground for applying the Lagrangian framework we will derive the Lagrangian kernels in Sec.\,\ref{sec:Lag}.
\paragraph*{Density power spectrum}
For the density power spectrum the effect of the coarse-grained fluid equations \eqref{Fluidcg} is simply to coarse-grain the power spectrum obtained from SPT according to
\begin{subequations}
\begin{align}
P_{\bar\delta\bar\delta}(k) = \bar P_{\delta\delta}(k)  = \exp\left(-\sigx^2 k^2\right) P_{\delta\delta}(k) \,. \label{PSdeltacg}
\end{align}
\end{subequations}
This result holds at any order in SPT and shows that, as expected, the smoothing becomes effective only at small scales $k \gtrsim 1/\sigx$. Note that, since the power spectrum is quadratic in $\delta$ it gets smoothed with $\sqrt 2 \sigx$ when $\delta$ is coarse-grained on scale $\sigx$. Therefore we will write in the following $\bar P(k) := \exp\left(-\sigx^2 k^2\right) P(k)$ even if $\bar \delta (\vk) := \exp\left(-\tfrac{1}{2}\sigx^2 k^2\right) \delta(\vk)$. The resulting power spectrum depicted in Fig.\,\ref{ddspectrum} shows that power on small spatial scales corresponding to large $k$ is suppressed due to the coarse-graining.

\paragraph*{Velocity power spectrum}
The effect of the coarse-graining onto the velocity power spectrum is more involved than for the density power spectrum since the cgSPT kernels for the mass-weighted velocity have to be evaluated according to Eq.\,\eqref{Vkernel}. At linear level the mass-weighted velocity kernel $\bar{\v{V}}_1$ corresponds to a smoothing of the microscopic one. This directly translates to the power spectrum of  the macroscopic velocity divergence $\bar\theta$ and the vorticity which vanishes identically
\begin{subequations}
\begin{align}
P_{\bar\theta\bar\theta,L}(k) & = \bar P_{\theta\theta,L}(k) \,, \\  
P_{\bar{\v{w}}\bar{\v{w}},L}(k) &= \bar P_{\v{w}\v{w},L}(k) = 0\,.
\end{align}
\end{subequations}
Note that since $\bar \theta_1(\vk) = - \bar\delta_1(\vk)$ the linear velocity power spectrum is identical to the linear density power spectrum when expressed in the same units.

\begin{figure}[h!]
\includegraphics[width=0.48\textwidth]{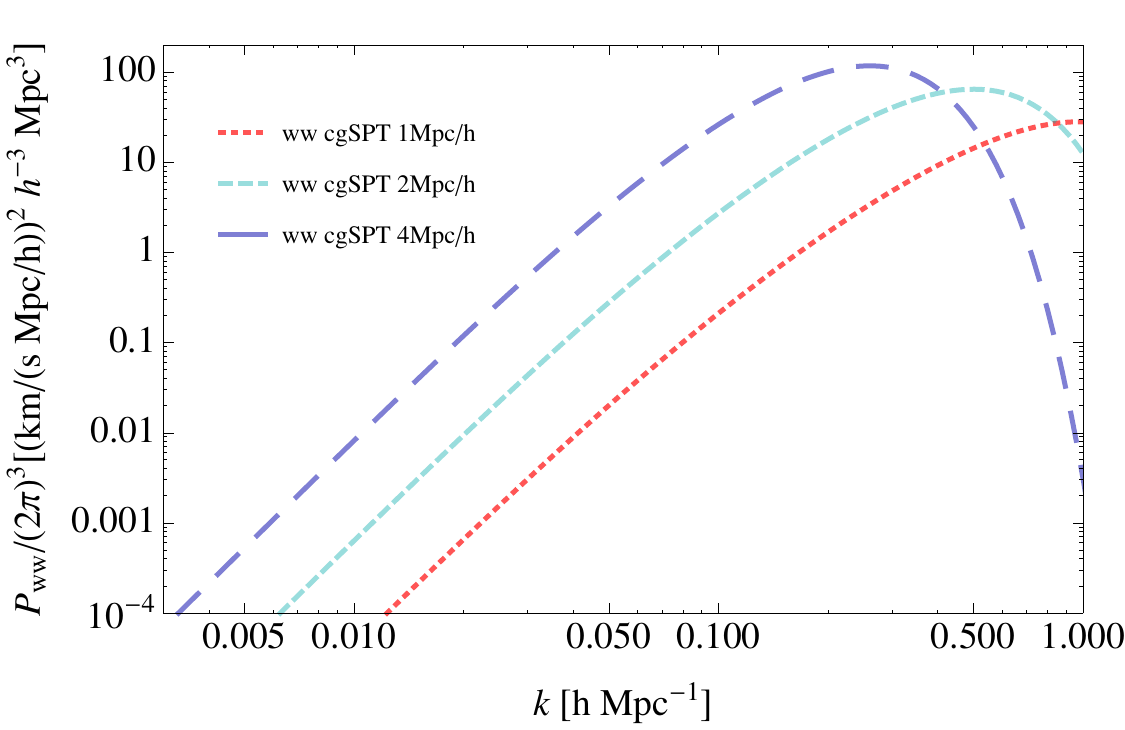}
\caption{Matter vorticity power spectrum for cgSPT in 1-loop order perturbation theory for different smoothing scales.}
\label{wwspectrum}
\end{figure}

At one-loop level the different contributions to the total velocity kernel $\bar{\v{V}}_n$ according to Eq.\,\eqref{Vkernel} have been evaluated explicitly in Appendix \ref{AppPower}. As can be seen in Fig.\,\ref{vvspectrum}, the effect of the dynamical coarse-graining (cgSPT) for the velocity $\v{v}$ power spectrum differs from coarse-graining the initial conditions in SPT (SPTcg). Most notably, our coarse-graining procedure determining the mass-weighted velocity $\bar{\v{v}}$ introduces a nonzero vorticity $\v{\bar w}=\vnabla \times \bar{\v{v}}$ which manifests itself from second order on and is shown in Fig.\,\ref{wwspectrum}. The vorticity marginally affects the velocity power spectrum at one-loop order via its contribution $P_{\bar {\v{w}}\bar{\v{w}},22}$. However, this contribution present in cgSPT is a fundamental difference to SPT where vorticity cannot be sourced when it is zero initially. The corresponding expression can be computed readily from the recursion relations \eqref{Vkernel} and reads
\begin{align}
\label{PSvort}
\notag P_{\bar{\v{w}}\bar{\v{w}},22}(k) &= 2\int \frac{\vol{3}{p}}{(2\pi)^3} \left|\bar{\v{W}}_2^{(s)}(\vp,\vk-\vp)\right|^2  P_L(p)P_L(|\vk-\vp|)  \\
\notag &= \frac{k^3}{2\pi^2} \int_0^\infty dr\ \int_{-1}^1 dx \ \bar P_L(kr) \bar P_L\left(k\sqrt{1-2rx+r^2}\right)   \\
 & \quad\ \times \frac{\left(1-x^2\right) (1-2 r x)^2 \left(e^{\sigx^2k^2 (r^2-rx)}-1\right)^2}{4 \left(r^2-2 r x+1\right)^2} \,.
\end{align}
Interestingly, the only effect of increasing the coarse-graining scale is to cause vorticity to become relevant at larger length scales whereas the shape of the vorticity power spectrum remains unchanged and the slope seems to be universal.
\begin{figure}[t!]
\includegraphics[width=0.48\textwidth]{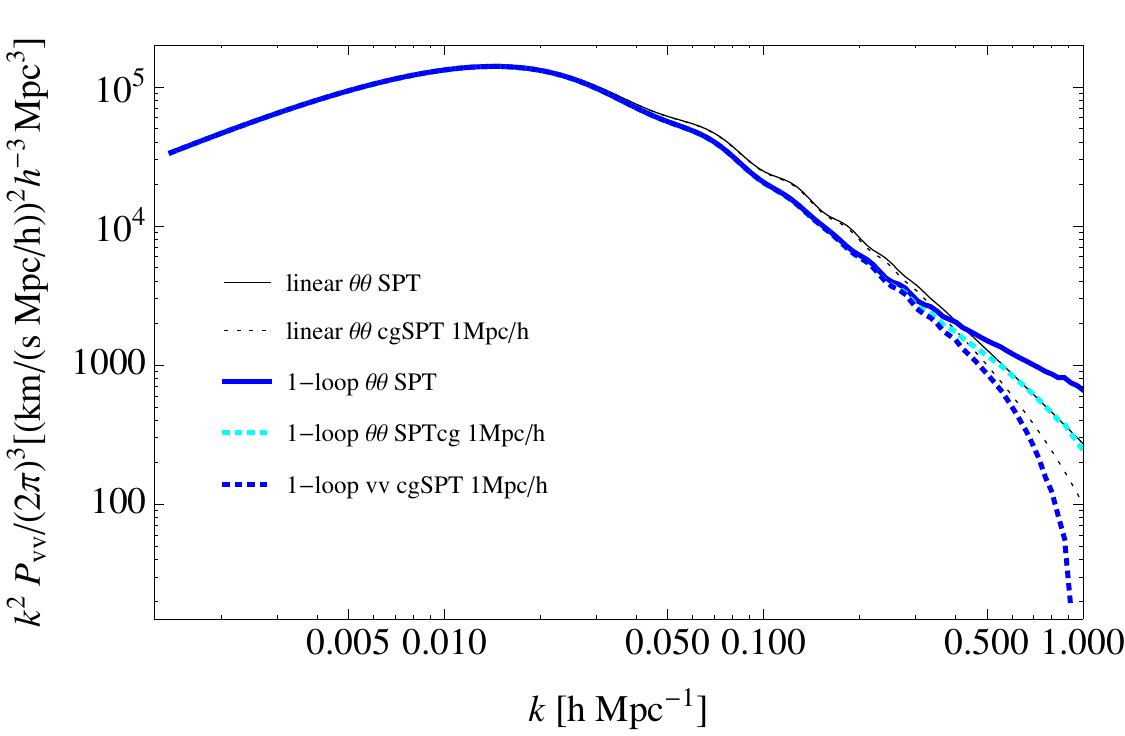}
\caption{Comparison between matter velocity power spectrum for SPT, cgSPT and SPTcg in 1st (lin) and 2nd (1-loop) order perturbation theory. }
\label{vvspectrum}
\end{figure}

\begin{figure}[t!]
\includegraphics[width=0.455\textwidth]{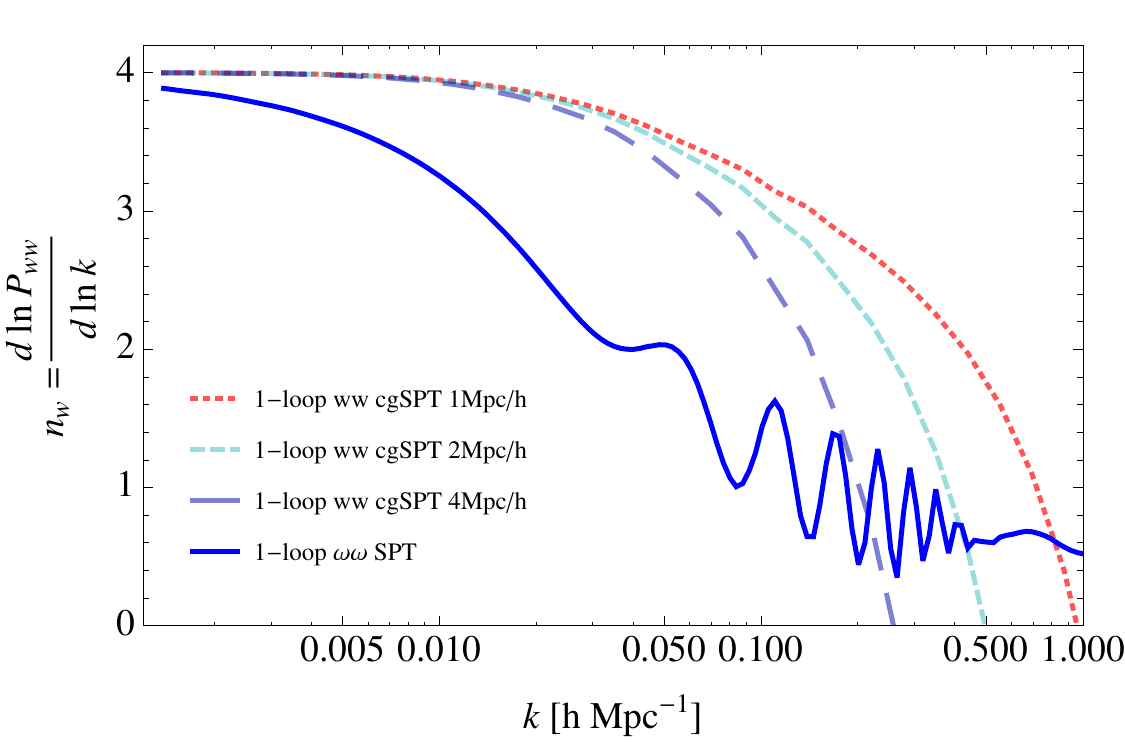}
\caption{Spectral index $n_{\v{w}}$ of the vorticity power spectrum $P_{\v{w}\v{w}} \simeq (k/k_{NL})^{n_{\v{w}}}$ as function of wavenumber $k$}
\label{wwspectrumslope}
\end{figure}

In \cite{S00,PS09} it has been suggested that the basic features of the vorticity power spectrum can be understood when assuming that the vorticity in regions which underwent shell-crossing is induced by mass-weighting the single-stream velocities. However, they used as estimate $\v{w} \sim \vnabla \times [(1+\delta)\v{v}]/(1+\delta)$ where, in contrast, no real mass-weighting has been performed. Instead, they considered the vorticity of the momentum $\v{j}=(1+\delta)\v{v}$ and afterwards divided by $(1+\delta)$. In turn, our approach based on coarse-graining automatically implements this idea correctly and yields a vorticity according to $\bar{\v{w}}=\vnabla \times \bar{\v{v}} = \vnabla \times \left[ \overline{(1+\delta)\v{v}}/(1+\bar \delta)\right]$. We want to emphasize that this vorticity is induced by the underlying smoothing scale and hence of purely macroscopic origin rather than related to microscopic shell-crossing. Any vorticity on small scales generated by multi-streaming effects would add to that large scale vorticity induced by coarse-graining the single-stream physics. However, as we will see shortly, there exists an optimal smoothing scale such that the neglected microscopic vorticity does only contribute on scales smaller than the smoothing scale which makes it possible to attribute the entire large scale vorticity to the coarse graining of a dust fluid completely free of any microscopic vorticity. Since in \cite{PS09} there was no prediction for the amplitude of the vorticity power spectrum we can only compare the spectral index $n_{\v{w}}:=d\ln P_{\v{w}\v{w}}/d\ln k$ which is depicted in Fig.\,\ref{wwspectrumslope}.

Our results agree with predictions made in EFT of LSS \cite{CFGS13}, which give $P_{\v{w}\v{w}} \propto (k/k_{\rm{NL}})^{n_{\v{w}}} $ with
\begin{align}
n_{\v{w}}=
\begin{cases}
\ 4& \text{for}\quad k\ [h\text{Mpc}^{-1} ] \lesssim 0.1 \\
\ 3.6& \text{for}\quad 0.1\lesssim k\ [h\text{Mpc}^{-1} ] \lesssim 0.3\\
\ 2.8& \text{for}\quad 0.3 \lesssim k\ [h\text{Mpc}^{-1} ] \lesssim 0.6 \\
\end{cases}\,,
\end{align}
where $k_{\rm{NL}}$ is the nonlinear scale in EFT of LSS \cite{CFGS13}. Furthermore we can clearly see that the spectral index of the vorticity caused by a mass-weighted velocity differs significantly from the estimation $\v{w}\sim \vnabla \times \left[(1+\delta)\v{v}\right]/(1+\delta)$ made in \cite{S00,PS09}, which is the solid blue wiggly line Fig.\,\ref{wwspectrumslope} denoted by $\omega \omega$.\footnote{At 2nd order this power spectrum is equivalent to the power spectrum of vector metric perturbations $\v{\omega}$ as calculated in Eq. 4.9c in \cite{KUH14}.}

\paragraph*{Cross spectrum}
The cross spectrum can be determined in the same way as the power spectra by employing the kernels for the density contrast and velocity divergence obtained before. At lowest order the cross spectrum of cgSPT is trivially related to the SPT cross spectrum
\begin{align}
P_{\bar\delta\bar\theta,L}(k) &= \bar P_{\delta\theta,L}(k)\,.
\end{align}
which is again just a scaled version of the density power spectrum. Beyond linear order there is another contribution besides the trivial smoothing which slightly affects the cross spectrum, namely the effect of the modified recursion relation for the kernel $\bar G_n$ of $\bar \theta_n$ according to \eqref{Vkernel} which differs from the smoothing of the kernel $G_n$ of $\theta_n$. The cross spectrum at 1-loop order, whose explicit expression can be found in Appendix \ref{AppPower}, is shown in Fig.\,\ref{dtspectrum}.

\paragraph*{Comparison to N-body simulations}

We can compare our analytical results to power and cross spectra obtained from cosmological numerical simulations, as given in \cite{PS09}, \cite{PW09} and \cite{H14}. Note that within both works a different Fourier convention has been employed. Therefore we show our power spectra divided by $(2\pi)^3$ in order to allow for comparison which reveals good qualitative agreement. Note also that our $\theta_n(k)$ is dimensionless because we factored out $\sH a^n$ in \eqref{pertexpvcg}, while $\theta$ is measured in km/(s Mpc/$h$) to facilitate comparison with \cite{H14} .

In \cite{PS09} it was noted that the vorticity power spectrum shows significant sensitivity on the mass resolution which was confirmed by \cite{H14}, compare Fig.\,3 in \cite{PS09} and Fig.\,12 in \cite{H14}. Similarly, a strong dependence on the smoothing scale shows up in the vorticity power spectrum for cgSPT, see Fig.\,\ref{vvspectrum}. Comparing the theoretical prediction for the vorticity power spectrum using different coarse-graining scales depicted in Fig.\,\ref{vvspectrum} with the converged spectrum Fig.\,12 in \cite{H14} we obtain the best agreement in amplitude for a smoothing scale of $\sigx\simeq 1\Mpc$. The main effect of an increasing coarse-graining scale is to shift the wavenumber at which vorticity becomes relevant to smaller values corresponding to larger length scales. However, the spectral index $n_w$ is a rather universal feature of the vorticity power spectrum and was determined in \cite{H14} as a function of $k$. Its asymptotic values were found to agree with $P_{\v{w}\v{w}}\propto k^{5/2}$ on large scales and $P_{\v{w}\v{w}}\propto k^{-3/2}$ on small scales. Due to the coarse-graining in our formalism small spatial scales corresponding to large $k>1/\sigx$ are not accessible. On intermediate scales we find reasonable agreement for the spectral index $n_{\v{w}}$ predicted by cgSPT, shown as dashed lines in Fig.\,\ref{wwspectrumslope}, with N-body simulations, compare Fig.\,14 in \cite{H14}. The spectral index $n_{\v{w}}$ obtained from estimating the vorticity according to \cite{S00,PS09} is qualitatively different, see the solid blue line in Fig.\,\ref{wwspectrumslope}.
\begin{figure}[t!]
\includegraphics[width=0.47\textwidth]{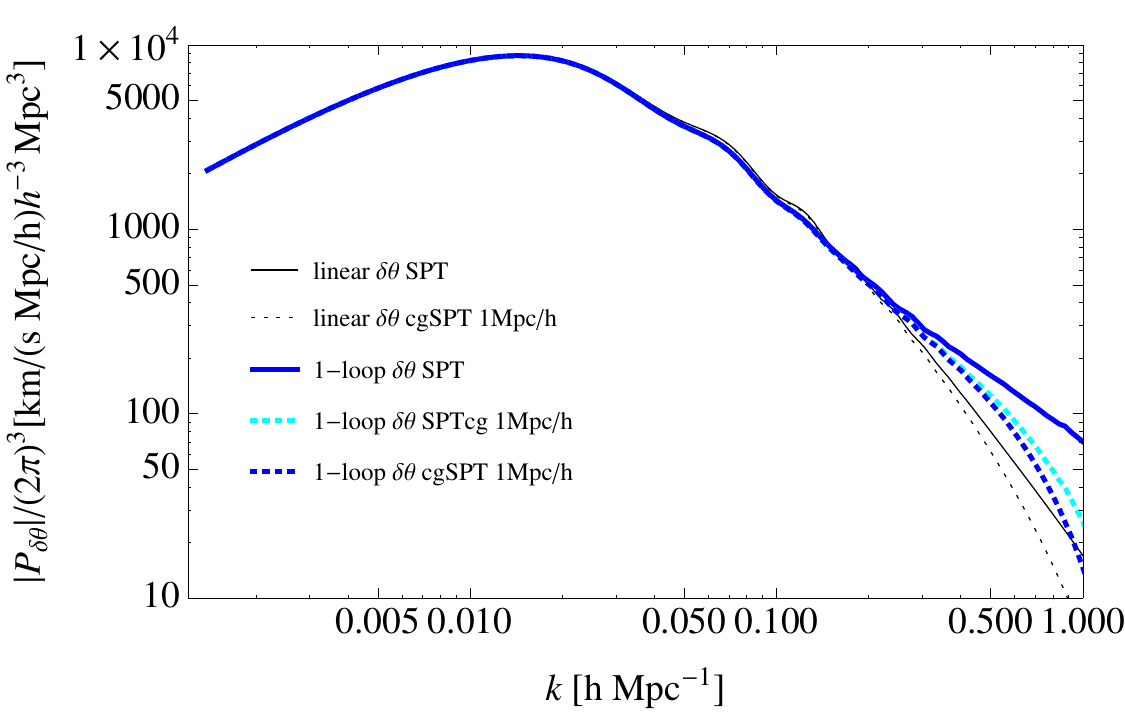}
\caption{Comparison between cross spectrum for SPT, cgSPT and SPTcg in 1st (lin) and 2nd (1-loop) order perturbation theory. }
\label{dtspectrum}
\end{figure}

\section{Lagrangian Perturbation Theory}
\label{sec:Lag}
Well before the onset of strong non-linearity and shell crossing the dynamics of a perfect pressureless fluid qualitatively resembles coarse-grained hydrodynamics. The Zel'dovich approximation based on the dust fluid has proven quite successful in the mildly nonlinear regime \cite{BMW94,Ta14}. Since it should retain its applicability in the coarse-grained dust model, we will consider the effect of the coarse-graining in Eulerian space onto LPT. This also paves the way to generalize the Post-Zel'dovich approximation or resummation schemes like iPT \cite{M11} and CLPT \cite{CRW13} to the coarse-grained dust model. 

We perform a transformation to Lagrangian coordinates $\v{q}$ in which the coarse-grained fluid positions are given by the old Eulerian coordinates $\v{x}= \v{q} + \bar{\v{\varPsi}}$, where the displacement field $\bar{\v{\varPsi}}(\v{q},\tau)$ is given as the integral lines of the coarse-grained Eulerian velocity emanating at $\v{q}$
\begin{align}
\bar{\v{v}}(\v{q},\tau) = \partial_{\tau}|_{q}\bar{\v{\varPsi}}(\v{q},\tau) \,. \label{defPsi}
\end{align}
Note that, clearly a coarse-graining in Eulerian space is not equivalent to a direct coarse-graining in Lagrangian space. $\bar{\v{\varPsi}}$ is defined via the coarse-grained Eulerian quantities $\bar\delta$ and $\bar{\v{v}}$ according to \eqref{defPsi} and not to be understood as the coarse-graining of ${\v{\varPsi}}$. Therefore, we will perform the mapping from SPT to LPT, proceeding along the lines of \cite{RB12}, to determine the perturbative kernels in Lagrangian space that correspond to the coarse-grained Eulerian kernels derived before.
When we define the Jacobian of the transformation as
\begin{align} \label{FijCG}
\bar F_{ij}=\frac{\del  x_i}{\del q_j} = \delta_{ij}+\bar \varPsi_{i,j}\,, \quad  J_{\bar F}=\det\left[\delta_{ij}+\bar{\varPsi}_{i,j}\right] \,,
\end{align}
we obtain the same relation between $\bar \delta $ and $\bar{\v{\varPsi}}$ as for dust 
\begin{align}
1+\bar \delta=J^{-1}_{\bar F} \,,\label{JFtoDeltaCG}
\end{align}
since $\bar \delta$ and $\bar{\v{v}}$ fulfill the continuity equation \eqref{Conticg}.
Therefore we have in Fourier space 
\begin{align}
\bar{\delta} (\vk) 
&= \int \vol{3}{q} e^{-i\vk\cdot\vq}  \left(e^{-i\vk\cdot\bar{\v{\varPsi}}(\vq)}-1\right) \,.\label{deltathetaFromPsiCG}
\end{align}
Next, we expand the displacement field $\v{\bar \varPsi}(\tau,\vk)$ perturbatively
\begin{align}
\label{pertExpPsiCG}
\bar{\v{\varPsi}}(\tau,\vq) &= \sum_{n=1}^\infty a^n(\tau)\bar{\v{\varPsi}}^{(n)}(\vq) \,,
\end{align}
and express the different orders $\v{\bar \varPsi}^{(n)}$ with the help of perturbative kernels $\v{\bar L}^{(n)}$ defined as
 \begin{align}
\label{defLCG}
\notag \bar{\v{\varPsi}}^{(n)}(\vk)&= i \int \frac{\vol{3}{p_1}\ldots\ \vol{3}{p_n}}{(2\pi)^{3(n-1)}} \delta_{\rm D}(\vk-\vp_{1\cdots n})  \bar{\v{L}}^{(n)}(\vp_1,\ldots,\vp_n) \times \\
&\qquad\qquad\qquad\qquad \times \delta_1(\vp_1) \cdots \delta_1(\vp_n) \,.
\end{align}
The kernel $\v{\bar L}^{(n)}= \v{\bar S}^{(n)}+ \v{\bar T}^{(n)}$ is split into its longitudinal $\v{\bar S}^{(n)}$ and transverse part $\v{\bar T}^{(n)}$ which fulfill $\vk \times  \v{\bar S}^{(n)}(\vp_1,\ldots,\vp_n) = 0$ and $\vk\cdot \v{\bar T}^{(n)}(\vp_1,\ldots,\vp_n)=0$, respectively.

The expressions for microscopic density $\delta$ and velocity divergence $\theta$ in terms of displacements $\v{\varPsi}$ given in \cite{RB12} can be directly translated to those between the macroscopic quantities $\bar \delta$ and $\bar \theta$ and $\bar{\v{\varPsi}}$. In addition we need a corresponding expression for the vorticity $\v{\bar w}$ which is present in the coarse-grained dust model but absent in the dust model.  Using the Jacobian $\bar F_{ij} = \del x_i/\del q_j$, we can write the vorticity as
\begin{subequations}
\begin{align}
\bar w_i&=(\v{\nabla}_{\! \! x}\times \v{\bar v})_i = \varepsilon_{ijk} \partial_{x_j} \bar v_k = \varepsilon_{ijk} (\bar F_{mj})^{-1} \bar F_{km}' \,.
\end{align}
By multiplying with $J_{\bar F} = \det{\bar F_{ij}}$ and inserting $\bar F_{ij}$ according to \eqref{FijCG} and using Eqs. (3d) and (6f) from \cite{EB97} we obtain
\begin{align}
\label{JW}
J_{\bar F} \bar w_i &= \bar F_{in} \varepsilon_{njk}\bar F_{lj}\bar F_{lk}'\\
\notag &= -\varepsilon_{ijk} \left(\bar \varPsi_{k,j}' -\bar \varPsi_{l,j} \bar \varPsi_{l,k}' \right) - \bar \varPsi_{i,n}\varepsilon_{njk}\left(\bar \varPsi_{k,j}' -\bar \varPsi_{l,j} \bar \varPsi_{l,k}' \right)\,.
\end{align}
This allows to express the vorticity $\v{\bar w}$ in Fourier space entirely in terms of $\v{\bar \varPsi}$ according to
\begin{align} \label{wFromPsiCG}
\bar w_i(\v{k}) 
&=\int \vol{3}{q} e^{-i \v{k}\cdot \v{q} - i \v{k}\cdot \v{\bar\varPsi}} J_{\bar F} \bar w_i(\v{q})\,.
\end{align}
\end{subequations}
Combining these results we can match the Eulerian and Lagrangian expressions at each order. Thereby we obtain expressions for the vorticity kernels $\v{\bar W}$ in terms of the Lagrangian kernels $\v{\bar S}$ and $\v{\bar T}$. For the density contrast we can proceed analogously to the dust case as described in \cite{RB12} to obtain those relations. The results expressing the Eulerian kernels of vorticity $\v{\bar W}$ and density $\bar F$ in terms of the longitudinal $\v{\bar S}$ and transverse parts $\v{\bar T}$ of the Lagrangian kernels can be found in App.\,\ref{AppLag}. Since $\v{k}\cdot\v{\bar T} =0$ we have $ \v{k}\times(\v{k}\times \v{\bar T}) = -k^2 \v{\bar T}$ which allows to invert the $\v{\bar W}$ relation \eqref{WSTrelation} for $\v{\bar T}$. Furthermore the $\bar F$ relation \eqref{FSTrelation} can be easily inverted for $\v{\bar S}$. 
Therefore, the longitudinal $\v{\bar S}^{(n)}$ and transverse $\v{\bar T}^{(n)}$ kernels of the displacement field are related to the Eulerian kernels for density $\bar F_n$ and vorticity $\bar{\v{W}}_n$ via

\begin{widetext}
 \begin{subequations}
 \label{SandT}
\begin{align}
\v{\bar S}^{(1)}(\v{p}_1)  &= \exp\left(-\tfrac{1}{2}\sigx^2 p_1^2\right)\frac{\v{p}_1}{p_1^2} \,, \qquad\qquad\qquad\qquad\qquad
\v{\bar T}^{(1)}(\v{p}_1)  = 0  \,,\label{ST1}\\
\v{\bar S}^{(2)}(\v{p}_1,\v{p}_2) &= \frac{\v{p}_{12}}{p_{12}^2}\left( \bar F_2^{\rm s}  -\frac{1}{2} \left(\v{p}_{12} \cdot \v{\bar S}^{(1)}\right)\ \left(\v{p}_{12} \cdot \v{\bar S}^{(1)}\right)\right) \,,\qquad
\v{\bar T}^{(2)}(\v{p}_1,\v{p}_2) =\frac{1}{2} \frac{1}{p_{12}^2}\v{p}_{12}\times \bar{\v{W}}^{(2)}_{\rm s}(\v{p}_1,\v{p}_2) \,,\label{ST2}\\
 \v{\bar S}^{(3)}(\v{p}_1,\v{p}_2,\v{p}_3)&= \frac{\v{p}_{123}}{p_{123}^2}\Bigg\{ \bar F_3^{\rm s} - \frac{1}{6} \left(\v{p}_{123} \cdot \v{\bar S}^{(1)}\right)\ \left(\v{p}_{123} \cdot \v{\bar S}^{(1)}\right)\ \left(\v{p}_{123} \cdot \v{\bar S}^{(1)}\right)- \frac{1}{3}\stackrel{+ \text{cyclic permutation of } (\vp_1,\vp_2,\vp_3)}{\left(\v{p}_{123}\cdot \v{\bar S}^{(1)}\right)\,\left(\v{p}_{123}\cdot \left[ \v{\bar S}^{(2)}+\v{\bar T}^{(2)}\right] \right) } \Bigg\} \label{ST3} \,,\\
\notag  \v{\bar T}^{(3)} (\v{p}_1,\v{p}_2,\v{p}_3)&=\frac{1}{3}\frac{\v{p}_{123}}{p_{123}^2} \times \Bigg\{   \bar{\v{W}}^{(3)}_{\rm s} + \frac{1}{3}\stackrel{+ \text{cyclic permutation of } (\vp_1,\vp_2,\vp_3)\qquad + \text{cyclic permutation of } (\vp_1,\vp_2,\vp_3) \qquad + \text{cyclic permutation of } (\vp_1,\vp_2,\vp_3)}{\Big[\v{p}_1\times \v{p}_{23} \left(\v{\bar S}^{(1)}\cdot \left[ \v{\bar S}^{(2)}+ \v{\bar T}^{(2)} \right]\right)+ 2\, \v{p}_{23}\times \v{\bar T}^{(2)} \left( \v{p}_{123}\cdot \v{\bar S}^{(1)}\right)-2\,\v{\bar S}^{(1)}\,\left(\v{p}_1\cdot\left[\v{p}_{23} \times \v{\bar T}^{(2)} \right]\right)\Big]}\Bigg\}\,.
\end{align}
\end{subequations}
For the sake of brevity we suppress the functional dependencies on the right hand side. They can be easily restored by attaching each kernel a dependence on $(\vp_{i}, \ldots, \vp_{i+n-1})$ in ascending order beginning with $i=1$ from the left, for example $\v{\bar S}^{(1)}\cdot\v{\bar T}^{(2)}:=\v{\bar S}^{(1)}(\vp_1)\cdot\v{\bar T}^{(2)}(\vp_2,\vp_3)$. We defined $ \bar{\v{W}}^{(n)}_{\rm s}(\v{p}_1,..,\v{p}_n) := 1/n! \sum_{\sigma\in S_n} \bar{\v{W}}^{(n)}(\v{p}_{\sigma(1)},..,\v{p}_{\sigma(n)} )\,,$ where the sum goes over all $n!$ permutations of $n$ indices. Note that our kernels correctly reproduce to the standard dust kernels, given in \cite{RB12} in the limit $\sigx\rightarrow 0$. 
\end{widetext}
\section{Conclusion and Outlook}
\label{sec:concl}
In order to model collisionless selfgravitating matter we considered a coarse-grained dust fluid, which is in turn a good model for cold dark matter in the single-stream regime and entire halos on large scales. We studied it perturbatively in the Eulerian frame and derived recursion relations Eqs.\,\eqref{cgrecursrel} for the Fourier kernels of the coarse-grained density contrast $\bar\delta$ and the mass-weighted velocity $\v{\bar v}$. Those recursive expressions are given in terms of the standard perturbation kernels for the microscopic density contrast $\delta$ and the velocity divergence $\theta$ of a pressureless (dust) fluid. \\
We computed the corresponding power and cross spectra of the coarse-grained density contrast $\bar\delta$ and the mass-weighted velocity $\v{\bar v}$ up to 1-loop order perturbation theory and compared them to the standard dust case. Our study revealed that in the coarse-grained dynamics a vorticity $\v{\bar w}=\vnabla\times\v{\bar v}$ is generated dynamically which becomes manifest already at 1-loop order in the power spectrum Eq.\,\eqref{PSvort}. The magnitude, shape and spectral index of the analytically predicted vorticity power spectrum, see Fig.\,\ref{vvspectrum}, exhibits qualitatively good agreement with recent measurements from N-body simulations \cite{H14}. 
This suggests that the large-scale vorticity observed in N-body simulations can be interpreted as a smoothing effect within a single-streaming dust fluid, rather than in terms of the actual microscopic physics, in which vorticity is generated by shell-crossing effects on small scales. The fact that the macroscopic vorticity $\bar{\v{w}}$ can be calculated from a single-streaming dust fluid, makes it accessible to perturbation theory, see Eqs. \eqref{wn}, \eqref{Vkernel}. A similar phenomenon has been observed before in \cite{P11} for velocity dispersion. In principle, comparing N-body measurements of velocity dispersion in addition to vorticity to the theoretical prediction in our framework opens up the possibility to fix the two parameters $\sigx$ and $\sigp$ involved in the coarse-grained dust model. Once the smoothing scales $\sigx$ and $\sigp$ are known, the functional form of all higher cumulants is determined unambiguously which would then allow to test the viability of the coarse-grained dust model. The idea of this procedure is in line with the EFT approach which becomes predictive once a limited number of effective parameters are determined from simulations or observations.
Interestingly, the scale dependence of the spectral index for vorticity shown in Fig.\,\ref{wwspectrumslope} resembles the observations made within the EFT of LSS \cite{CFGS13}. 

Finally, we explained how the perturbation kernels of the displacement field $\v{\bar\varPsi}$ in the Lagrangian framework can be obtained from the Eulerian kernels of the density contrast $\bar\delta$ and the vorticity $\v{\bar w}$. To pave the way for applications in the context of LPT we gave explicit expressions for the displacement kernels up to third order in Eqs.\,\eqref{SandT}. Based on these results we will determine the impact of the coarse-graining onto the correlation function of halos by extending CLPT \cite{CRW13}, an approximation to the Post-Zel’dovich approximation, in a further investigation. Therein we will especially study how redshift space distortions affect the correlation function by means of the Gaussian streaming model \cite{RW11} and its possible extensions.
 
\section*{Acknowledgement}
The work of MK \& CU was supported by the DFG cluster of excellence ``Origin and Structure of the Universe''.  

\newcommand{\apjl}{Astrophys. J. Letters}
\newcommand{\apjs}{Astrophys. J. Suppl. Ser.}
\newcommand{\mnras}{Mon. Not. R. Astron. Soc.}
\newcommand{\pasj}{Publ. Astron. Soc. Japan}
\newcommand{\apss}{Astrophys. Space Sci.}
\newcommand{\aap}{Astron. Astrophys.}
\newcommand{\physrep}{Phys. Rep.}
\newcommand{\mpla}{Mod. Phys. Lett. A}
\newcommand{\jcap}{J. Cosmol. Astropart. Phys.}
\bibliography{cgDustbib}

\begin{thebibliography}{38}%
\makeatletter
\providecommand \@ifxundefined [1]{%
 \@ifx{#1\undefined}
}%
\providecommand \@ifnum [1]{%
 \ifnum #1\expandafter \@firstoftwo
 \else \expandafter \@secondoftwo
 \fi
}%
\providecommand \@ifx [1]{%
 \ifx #1\expandafter \@firstoftwo
 \else \expandafter \@secondoftwo
 \fi
}%
\providecommand \natexlab [1]{#1}%
\providecommand \enquote  [1]{``#1''}%
\providecommand \bibnamefont  [1]{#1}%
\providecommand \bibfnamefont [1]{#1}%
\providecommand \citenamefont [1]{#1}%
\providecommand \href@noop [0]{\@secondoftwo}%
\providecommand \href [0]{\begingroup \@sanitize@url \@href}%
\providecommand \@href[1]{\@@startlink{#1}\@@href}%
\providecommand \@@href[1]{\endgroup#1\@@endlink}%
\providecommand \@sanitize@url [0]{\catcode `\\12\catcode `\$12\catcode
  `\&12\catcode `\#12\catcode `\^12\catcode `\_12\catcode `\%12\relax}%
\providecommand \@@startlink[1]{}%
\providecommand \@@endlink[0]{}%
\providecommand \url  [0]{\begingroup\@sanitize@url \@url }%
\providecommand \@url [1]{\endgroup\@href {#1}{\urlprefix }}%
\providecommand \urlprefix  [0]{URL }%
\providecommand \Eprint [0]{\href }%
\providecommand \doibase [0]{http://dx.doi.org/}%
\providecommand \selectlanguage [0]{\@gobble}%
\providecommand \bibinfo  [0]{\@secondoftwo}%
\providecommand \bibfield  [0]{\@secondoftwo}%
\providecommand \translation [1]{[#1]}%
\providecommand \BibitemOpen [0]{}%
\providecommand \bibitemStop [0]{}%
\providecommand \bibitemNoStop [0]{.\EOS\space}%
\providecommand \EOS [0]{\spacefactor3000\relax}%
\providecommand \BibitemShut  [1]{\csname bibitem#1\endcsname}%
\let\auto@bib@innerbib\@empty
\bibitem [{\citenamefont {{Peebles}}(1980)}]{P80}%
  \BibitemOpen
  \bibfield  {author} {\bibinfo {author} {\bibfnamefont {P.~J.~E.}\
  \bibnamefont {{Peebles}}},\ }\href@noop {} {\emph {\bibinfo {title} {{The
  large-scale structure of the universe}}}}\ (\bibinfo {year}
  {1980})\BibitemShut {NoStop}%
\bibitem [{\citenamefont {{Bernardeau}}\ \emph {et~al.}(2002)\citenamefont
  {{Bernardeau}}, \citenamefont {{Colombi}}, \citenamefont {{Gazta{\~n}aga}},\
  and\ \citenamefont {{Scoccimarro}}}]{B02}%
  \BibitemOpen
  \bibfield  {author} {\bibinfo {author} {\bibfnamefont {F.}~\bibnamefont
  {{Bernardeau}}}, \bibinfo {author} {\bibfnamefont {S.}~\bibnamefont
  {{Colombi}}}, \bibinfo {author} {\bibfnamefont {E.}~\bibnamefont
  {{Gazta{\~n}aga}}}, \ and\ \bibinfo {author} {\bibfnamefont {R.}~\bibnamefont
  {{Scoccimarro}}},\ }\href {\doibase 10.1016/S0370-1573(02)00135-7} {\bibfield
   {journal} {\bibinfo  {journal} {\physrep}\ }\textbf {\bibinfo {volume}
  {367}},\ \bibinfo {pages} {1} (\bibinfo {year} {2002})},\ \Eprint
  {http://arxiv.org/abs/astro-ph/0112551} {astro-ph/0112551} \BibitemShut
  {NoStop}%
\bibitem [{\citenamefont {{Ehlers}}\ and\ \citenamefont
  {{Buchert}}(1997)}]{EB97}%
  \BibitemOpen
  \bibfield  {author} {\bibinfo {author} {\bibfnamefont {J.}~\bibnamefont
  {{Ehlers}}}\ and\ \bibinfo {author} {\bibfnamefont {T.}~\bibnamefont
  {{Buchert}}},\ }\href {\doibase 10.1023/A:1018885922682} {\bibfield
  {journal} {\bibinfo  {journal} {General Relativity and Gravitation}\ }\textbf
  {\bibinfo {volume} {29}},\ \bibinfo {pages} {733} (\bibinfo {year} {1997})},\
  \Eprint {http://arxiv.org/abs/arXiv:astro-ph/9609036}
  {arXiv:astro-ph/9609036} \BibitemShut {NoStop}%
\bibitem [{\citenamefont {{Crocce}}\ and\ \citenamefont
  {{Scoccimarro}}(2006)}]{CS06}%
  \BibitemOpen
  \bibfield  {author} {\bibinfo {author} {\bibfnamefont {M.}~\bibnamefont
  {{Crocce}}}\ and\ \bibinfo {author} {\bibfnamefont {R.}~\bibnamefont
  {{Scoccimarro}}},\ }\href {\doibase 10.1103/PhysRevD.73.063519} {\bibfield
  {journal} {\bibinfo  {journal} {\prd}\ }\textbf {\bibinfo {volume} {73}},\
  \bibinfo {eid} {063519} (\bibinfo {year} {2006})},\ \Eprint
  {http://arxiv.org/abs/astro-ph/0509418} {astro-ph/0509418} \BibitemShut
  {NoStop}%
\bibitem [{\citenamefont {Matarrese}\ and\ \citenamefont
  {Pietroni}(2007)}]{MP07}%
  \BibitemOpen
  \bibfield  {author} {\bibinfo {author} {\bibfnamefont {S.}~\bibnamefont
  {Matarrese}}\ and\ \bibinfo {author} {\bibfnamefont {M.}~\bibnamefont
  {Pietroni}},\ }\href {\doibase 10.1088/1475-7516/2007/06/026} {\bibfield
  {journal} {\bibinfo  {journal} {JCAP}\ }\textbf {\bibinfo {volume} {0706}},\
  \bibinfo {pages} {026} (\bibinfo {year} {2007})},\ \Eprint
  {http://arxiv.org/abs/astro-ph/0703563} {arXiv:astro-ph/0703563 [astro-ph]}
  \BibitemShut {NoStop}%
\bibitem [{\citenamefont {Matsubara}(2008)}]{M08}%
  \BibitemOpen
  \bibfield  {author} {\bibinfo {author} {\bibfnamefont {T.}~\bibnamefont
  {Matsubara}},\ }\href {\doibase 10.1103/PhysRevD.78.109901,
  10.1103/PhysRevD.78.083519} {\bibfield  {journal} {\bibinfo  {journal}
  {Phys.Rev.}\ }\textbf {\bibinfo {volume} {D78}},\ \bibinfo {pages} {083519}
  (\bibinfo {year} {2008})},\ \Eprint {http://arxiv.org/abs/0807.1733}
  {arXiv:0807.1733 [astro-ph]} \BibitemShut {NoStop}%
\bibitem [{\citenamefont {Pueblas}\ and\ \citenamefont
  {Scoccimarro}(2009)}]{PS09}%
  \BibitemOpen
  \bibfield  {author} {\bibinfo {author} {\bibfnamefont {S.}~\bibnamefont
  {Pueblas}}\ and\ \bibinfo {author} {\bibfnamefont {R.}~\bibnamefont
  {Scoccimarro}},\ }\href {\doibase 10.1103/PhysRevD.80.043504} {\bibfield
  {journal} {\bibinfo  {journal} {Phys.Rev.}\ }\textbf {\bibinfo {volume}
  {D80}},\ \bibinfo {pages} {043504} (\bibinfo {year} {2009})},\ \Eprint
  {http://arxiv.org/abs/0809.4606} {arXiv:0809.4606 [astro-ph]} \BibitemShut
  {NoStop}%
\bibitem [{\citenamefont {Pietroni}\ \emph {et~al.}(2012)\citenamefont
  {Pietroni}, \citenamefont {Mangano}, \citenamefont {Saviano},\ and\
  \citenamefont {Viel}}]{P11}%
  \BibitemOpen
  \bibfield  {author} {\bibinfo {author} {\bibfnamefont {M.}~\bibnamefont
  {Pietroni}}, \bibinfo {author} {\bibfnamefont {G.}~\bibnamefont {Mangano}},
  \bibinfo {author} {\bibfnamefont {N.}~\bibnamefont {Saviano}}, \ and\
  \bibinfo {author} {\bibfnamefont {M.}~\bibnamefont {Viel}},\ }\href {\doibase
  10.1088/1475-7516/2012/01/019} {\bibfield  {journal} {\bibinfo  {journal}
  {JCAP}\ }\textbf {\bibinfo {volume} {1201}},\ \bibinfo {pages} {019}
  (\bibinfo {year} {2012})},\ \Eprint {http://arxiv.org/abs/1108.5203}
  {arXiv:1108.5203 [astro-ph.CO]} \BibitemShut {NoStop}%
\bibitem [{\citenamefont {{Hertzberg}}(2014)}]{H12}%
  \BibitemOpen
  \bibfield  {author} {\bibinfo {author} {\bibfnamefont {M.~P.}\ \bibnamefont
  {{Hertzberg}}},\ }\href {\doibase 10.1103/PhysRevD.89.043521} {\bibfield
  {journal} {\bibinfo  {journal} {\prd}\ }\textbf {\bibinfo {volume} {89}},\
  \bibinfo {eid} {043521} (\bibinfo {year} {2014})},\ \Eprint
  {http://arxiv.org/abs/1208.0839} {arXiv:1208.0839} \BibitemShut {NoStop}%
\bibitem [{\citenamefont {Carrasco}\ \emph {et~al.}(2012)\citenamefont
  {Carrasco}, \citenamefont {Hertzberg},\ and\ \citenamefont
  {Senatore}}]{CHS12}%
  \BibitemOpen
  \bibfield  {author} {\bibinfo {author} {\bibfnamefont {J.~J.~M.}\
  \bibnamefont {Carrasco}}, \bibinfo {author} {\bibfnamefont {M.~P.}\
  \bibnamefont {Hertzberg}}, \ and\ \bibinfo {author} {\bibfnamefont
  {L.}~\bibnamefont {Senatore}},\ }\href {\doibase 10.1007/JHEP09(2012)082}
  {\bibfield  {journal} {\bibinfo  {journal} {JHEP}\ }\textbf {\bibinfo
  {volume} {1209}},\ \bibinfo {pages} {082} (\bibinfo {year} {2012})},\ \Eprint
  {http://arxiv.org/abs/1206.2926} {arXiv:1206.2926 [astro-ph.CO]} \BibitemShut
  {NoStop}%
\bibitem [{\citenamefont {Carrasco}\ \emph {et~al.}(2013)\citenamefont
  {Carrasco}, \citenamefont {Foreman}, \citenamefont {Green},\ and\
  \citenamefont {Senatore}}]{CFGS13}%
  \BibitemOpen
  \bibfield  {author} {\bibinfo {author} {\bibfnamefont {J.~J.~M.}\
  \bibnamefont {Carrasco}}, \bibinfo {author} {\bibfnamefont {S.}~\bibnamefont
  {Foreman}}, \bibinfo {author} {\bibfnamefont {D.}~\bibnamefont {Green}}, \
  and\ \bibinfo {author} {\bibfnamefont {L.}~\bibnamefont {Senatore}},\
  }\href@noop {} {\  (\bibinfo {year} {2013})},\ \Eprint
  {http://arxiv.org/abs/1310.0464} {arXiv:1310.0464 [astro-ph.CO]} \BibitemShut
  {NoStop}%
\bibitem [{\citenamefont {Mercolli}\ and\ \citenamefont {Pajer}(2013)}]{M13}%
  \BibitemOpen
  \bibfield  {author} {\bibinfo {author} {\bibfnamefont {L.}~\bibnamefont
  {Mercolli}}\ and\ \bibinfo {author} {\bibfnamefont {E.}~\bibnamefont
  {Pajer}},\ }\href@noop {} {\  (\bibinfo {year} {2013})},\ \Eprint
  {http://arxiv.org/abs/1307.3220} {arXiv:1307.3220 [astro-ph.CO]} \BibitemShut
  {NoStop}%
\bibitem [{\citenamefont {{Carroll}}\ \emph {et~al.}(2013)\citenamefont
  {{Carroll}}, \citenamefont {{Leichenauer}},\ and\ \citenamefont
  {{Pollack}}}]{CLP13}%
  \BibitemOpen
  \bibfield  {author} {\bibinfo {author} {\bibfnamefont {S.~M.}\ \bibnamefont
  {{Carroll}}}, \bibinfo {author} {\bibfnamefont {S.}~\bibnamefont
  {{Leichenauer}}}, \ and\ \bibinfo {author} {\bibfnamefont {J.}~\bibnamefont
  {{Pollack}}},\ }\href@noop {} {\bibfield  {journal} {\bibinfo  {journal}
  {ArXiv e-prints}\ } (\bibinfo {year} {2013})},\ \Eprint
  {http://arxiv.org/abs/1310.2920} {arXiv:1310.2920 [hep-th]} \BibitemShut
  {NoStop}%
\bibitem [{\citenamefont {Manzotti}\ \emph {et~al.}(2014)\citenamefont
  {Manzotti}, \citenamefont {Peloso}, \citenamefont {Pietroni}, \citenamefont
  {Viel},\ and\ \citenamefont {Villaescusa-Navarro}}]{M14}%
  \BibitemOpen
  \bibfield  {author} {\bibinfo {author} {\bibfnamefont {A.}~\bibnamefont
  {Manzotti}}, \bibinfo {author} {\bibfnamefont {M.}~\bibnamefont {Peloso}},
  \bibinfo {author} {\bibfnamefont {M.}~\bibnamefont {Pietroni}}, \bibinfo
  {author} {\bibfnamefont {M.}~\bibnamefont {Viel}}, \ and\ \bibinfo {author}
  {\bibfnamefont {F.}~\bibnamefont {Villaescusa-Navarro}},\ }\href@noop {} {\
  (\bibinfo {year} {2014})},\ \Eprint {http://arxiv.org/abs/1407.1342}
  {arXiv:1407.1342 [astro-ph.CO]} \BibitemShut {NoStop}%
\bibitem [{\citenamefont {{Porto}}\ \emph {et~al.}(2013)\citenamefont
  {{Porto}}, \citenamefont {{Senatore}},\ and\ \citenamefont
  {{Zaldarriaga}}}]{PSZ13}%
  \BibitemOpen
  \bibfield  {author} {\bibinfo {author} {\bibfnamefont {R.~A.}\ \bibnamefont
  {{Porto}}}, \bibinfo {author} {\bibfnamefont {L.}~\bibnamefont {{Senatore}}},
  \ and\ \bibinfo {author} {\bibfnamefont {M.}~\bibnamefont {{Zaldarriaga}}},\
  }\href@noop {} {\bibfield  {journal} {\bibinfo  {journal} {ArXiv e-prints}\ }
  (\bibinfo {year} {2013})},\ \Eprint {http://arxiv.org/abs/1311.2168}
  {arXiv:1311.2168 [astro-ph.CO]} \BibitemShut {NoStop}%
\bibitem [{\citenamefont {Senatore}\ and\ \citenamefont
  {Zaldarriaga}(2014)}]{SZ14}%
  \BibitemOpen
  \bibfield  {author} {\bibinfo {author} {\bibfnamefont {L.}~\bibnamefont
  {Senatore}}\ and\ \bibinfo {author} {\bibfnamefont {M.}~\bibnamefont
  {Zaldarriaga}},\ }\href@noop {} {\  (\bibinfo {year} {2014})},\ \Eprint
  {http://arxiv.org/abs/1404.5954} {arXiv:1404.5954 [astro-ph.CO]} \BibitemShut
  {NoStop}%
\bibitem [{\citenamefont {Dominguez}(2000)}]{D00}%
  \BibitemOpen
  \bibfield  {author} {\bibinfo {author} {\bibfnamefont {A.}~\bibnamefont
  {Dominguez}},\ }\href {\doibase 10.1103/PhysRevD.62.103501} {\bibfield
  {journal} {\bibinfo  {journal} {Phys.Rev.}\ }\textbf {\bibinfo {volume}
  {D62}},\ \bibinfo {pages} {103501} (\bibinfo {year} {2000})}\BibitemShut
  {NoStop}%
\bibitem [{\citenamefont {{Buchert}}\ and\ \citenamefont
  {{Dom{\'{\i}}nguez}}(2005)}]{BD05}%
  \BibitemOpen
  \bibfield  {author} {\bibinfo {author} {\bibfnamefont {T.}~\bibnamefont
  {{Buchert}}}\ and\ \bibinfo {author} {\bibfnamefont {A.}~\bibnamefont
  {{Dom{\'{\i}}nguez}}},\ }\href {\doibase 10.1051/0004-6361:20052885}
  {\bibfield  {journal} {\bibinfo  {journal} {\aap}\ }\textbf {\bibinfo
  {volume} {438}},\ \bibinfo {pages} {443} (\bibinfo {year} {2005})},\ \Eprint
  {http://arxiv.org/abs/astro-ph/0502318} {astro-ph/0502318} \BibitemShut
  {NoStop}%
\bibitem [{\citenamefont {{Pietroni}}\ \emph {et~al.}(2012)\citenamefont
  {{Pietroni}}, \citenamefont {{Mangano}}, \citenamefont {{Saviano}},\ and\
  \citenamefont {{Viel}}}]{P12}%
  \BibitemOpen
  \bibfield  {author} {\bibinfo {author} {\bibfnamefont {M.}~\bibnamefont
  {{Pietroni}}}, \bibinfo {author} {\bibfnamefont {G.}~\bibnamefont
  {{Mangano}}}, \bibinfo {author} {\bibfnamefont {N.}~\bibnamefont
  {{Saviano}}}, \ and\ \bibinfo {author} {\bibfnamefont {M.}~\bibnamefont
  {{Viel}}},\ }\href {\doibase 10.1088/1475-7516/2012/01/019} {\bibfield
  {journal} {\bibinfo  {journal} {\jcap}\ }\textbf {\bibinfo {volume} {1}},\
  \bibinfo {eid} {019} (\bibinfo {year} {2012})},\ \Eprint
  {http://arxiv.org/abs/1108.5203} {arXiv:1108.5203 [astro-ph.CO]} \BibitemShut
  {NoStop}%
\bibitem [{\citenamefont {{Melott}}\ \emph {et~al.}(1994)\citenamefont
  {{Melott}}, \citenamefont {{Pellman}},\ and\ \citenamefont
  {{Shandarin}}}]{M94}%
  \BibitemOpen
  \bibfield  {author} {\bibinfo {author} {\bibfnamefont {A.~L.}\ \bibnamefont
  {{Melott}}}, \bibinfo {author} {\bibfnamefont {T.~F.}\ \bibnamefont
  {{Pellman}}}, \ and\ \bibinfo {author} {\bibfnamefont {S.~F.}\ \bibnamefont
  {{Shandarin}}},\ }\href@noop {} {\bibfield  {journal} {\bibinfo  {journal}
  {\mnras}\ }\textbf {\bibinfo {volume} {269}},\ \bibinfo {pages} {626}
  (\bibinfo {year} {1994})},\ \Eprint {http://arxiv.org/abs/astro-ph/9312044}
  {astro-ph/9312044} \BibitemShut {NoStop}%
\bibitem [{\citenamefont {{Coles}}\ \emph {et~al.}(1993)\citenamefont
  {{Coles}}, \citenamefont {{Melott}},\ and\ \citenamefont
  {{Shandarin}}}]{C93}%
  \BibitemOpen
  \bibfield  {author} {\bibinfo {author} {\bibfnamefont {P.}~\bibnamefont
  {{Coles}}}, \bibinfo {author} {\bibfnamefont {A.~L.}\ \bibnamefont
  {{Melott}}}, \ and\ \bibinfo {author} {\bibfnamefont {S.~F.}\ \bibnamefont
  {{Shandarin}}},\ }\href@noop {} {\bibfield  {journal} {\bibinfo  {journal}
  {\mnras}\ }\textbf {\bibinfo {volume} {260}},\ \bibinfo {pages} {765}
  (\bibinfo {year} {1993})}\BibitemShut {NoStop}%
\bibitem [{\citenamefont {Uhlemann}\ \emph {et~al.}(2014)\citenamefont
  {Uhlemann}, \citenamefont {Kopp},\ and\ \citenamefont {Haugg}}]{UKH14}%
  \BibitemOpen
  \bibfield  {author} {\bibinfo {author} {\bibfnamefont {C.}~\bibnamefont
  {Uhlemann}}, \bibinfo {author} {\bibfnamefont {M.}~\bibnamefont {Kopp}}, \
  and\ \bibinfo {author} {\bibfnamefont {T.}~\bibnamefont {Haugg}},\
  }\href@noop {} {\  (\bibinfo {year} {2014})},\ \Eprint
  {http://arxiv.org/abs/1403.5567} {arXiv:1403.5567 [astro-ph.CO]} \BibitemShut
  {NoStop}%
\bibitem [{\citenamefont {Hahn}\ \emph {et~al.}(2014)\citenamefont {Hahn},
  \citenamefont {Angulo},\ and\ \citenamefont {Abel}}]{H14}%
  \BibitemOpen
  \bibfield  {author} {\bibinfo {author} {\bibfnamefont {O.}~\bibnamefont
  {Hahn}}, \bibinfo {author} {\bibfnamefont {R.~E.}\ \bibnamefont {Angulo}}, \
  and\ \bibinfo {author} {\bibfnamefont {T.}~\bibnamefont {Abel}},\ }\href@noop
  {} {\  (\bibinfo {year} {2014})},\ \Eprint {http://arxiv.org/abs/1404.2280}
  {arXiv:1404.2280 [astro-ph.CO]} \BibitemShut {NoStop}%
\bibitem [{\citenamefont {Shandarin}\ and\ \citenamefont
  {Zeldovich}(1989)}]{SZ89}%
  \BibitemOpen
  \bibfield  {author} {\bibinfo {author} {\bibfnamefont {S.~F.}\ \bibnamefont
  {Shandarin}}\ and\ \bibinfo {author} {\bibfnamefont {Y.~B.}\ \bibnamefont
  {Zeldovich}},\ }\href {\doibase 10.1103/RevModPhys.61.185} {\bibfield
  {journal} {\bibinfo  {journal} {Rev. Mod. Phys.}\ }\textbf {\bibinfo {volume}
  {61}},\ \bibinfo {pages} {185} (\bibinfo {year} {1989})}\BibitemShut
  {NoStop}%
\bibitem [{\citenamefont {Baumann}\ \emph {et~al.}(2012)\citenamefont
  {Baumann}, \citenamefont {Nicolis}, \citenamefont {Senatore},\ and\
  \citenamefont {Zaldarriaga}}]{B10}%
  \BibitemOpen
  \bibfield  {author} {\bibinfo {author} {\bibfnamefont {D.}~\bibnamefont
  {Baumann}}, \bibinfo {author} {\bibfnamefont {A.}~\bibnamefont {Nicolis}},
  \bibinfo {author} {\bibfnamefont {L.}~\bibnamefont {Senatore}}, \ and\
  \bibinfo {author} {\bibfnamefont {M.}~\bibnamefont {Zaldarriaga}},\ }\href
  {\doibase 10.1088/1475-7516/2012/07/051} {\bibfield  {journal} {\bibinfo
  {journal} {JCAP}\ }\textbf {\bibinfo {volume} {1207}},\ \bibinfo {pages}
  {051} (\bibinfo {year} {2012})},\ \Eprint {http://arxiv.org/abs/1004.2488}
  {arXiv:1004.2488 [astro-ph.CO]} \BibitemShut {NoStop}%
\bibitem [{\citenamefont {{Jain}}\ and\ \citenamefont
  {{Bertschinger}}(1994)}]{JB94}%
  \BibitemOpen
  \bibfield  {author} {\bibinfo {author} {\bibfnamefont {B.}~\bibnamefont
  {{Jain}}}\ and\ \bibinfo {author} {\bibfnamefont {E.}~\bibnamefont
  {{Bertschinger}}},\ }\href {\doibase 10.1086/174502} {\bibfield  {journal}
  {\bibinfo  {journal} {\apj}\ }\textbf {\bibinfo {volume} {431}},\ \bibinfo
  {pages} {495} (\bibinfo {year} {1994})},\ \Eprint
  {http://arxiv.org/abs/arXiv:astro-ph/9311070} {arXiv:astro-ph/9311070}
  \BibitemShut {NoStop}%
\bibitem [{\citenamefont {Blas}\ \emph {et~al.}(2013)\citenamefont {Blas},
  \citenamefont {Garny},\ and\ \citenamefont {Konstandin}}]{BGK13}%
  \BibitemOpen
  \bibfield  {author} {\bibinfo {author} {\bibfnamefont {D.}~\bibnamefont
  {Blas}}, \bibinfo {author} {\bibfnamefont {M.}~\bibnamefont {Garny}}, \ and\
  \bibinfo {author} {\bibfnamefont {T.}~\bibnamefont {Konstandin}},\
  }\href@noop {} {\  (\bibinfo {year} {2013})},\ \Eprint
  {http://arxiv.org/abs/1309.3308} {arXiv:1309.3308 [astro-ph.CO]} \BibitemShut
  {NoStop}%
\bibitem [{\citenamefont {{Tassev}}(2014)}]{Ta14}%
  \BibitemOpen
  \bibfield  {author} {\bibinfo {author} {\bibfnamefont {S.}~\bibnamefont
  {{Tassev}}},\ }\href {\doibase 10.1088/1475-7516/2014/06/008} {\bibfield
  {journal} {\bibinfo  {journal} {\jcap}\ }\textbf {\bibinfo {volume} {6}},\
  \bibinfo {eid} {008} (\bibinfo {year} {2014})},\ \Eprint
  {http://arxiv.org/abs/1311.4884} {arXiv:1311.4884 [astro-ph.CO]} \BibitemShut
  {NoStop}%
\bibitem [{\citenamefont {Scoccimarro}(2000)}]{S00}%
  \BibitemOpen
  \bibfield  {author} {\bibinfo {author} {\bibfnamefont {R.}~\bibnamefont
  {Scoccimarro}},\ }\href@noop {} {\  (\bibinfo {year} {2000})},\ \Eprint
  {http://arxiv.org/abs/astro-ph/0008277} {arXiv:astro-ph/0008277 [astro-ph]}
  \BibitemShut {NoStop}%
\bibitem [{\citenamefont {{Kopp}}\ \emph {et~al.}(2014)\citenamefont {{Kopp}},
  \citenamefont {{Uhlemann}},\ and\ \citenamefont {{Haugg}}}]{KUH14}%
  \BibitemOpen
  \bibfield  {author} {\bibinfo {author} {\bibfnamefont {M.}~\bibnamefont
  {{Kopp}}}, \bibinfo {author} {\bibfnamefont {C.}~\bibnamefont {{Uhlemann}}},
  \ and\ \bibinfo {author} {\bibfnamefont {T.}~\bibnamefont {{Haugg}}},\ }\href
  {\doibase 10.1088/1475-7516/2014/03/018} {\bibfield  {journal} {\bibinfo
  {journal} {\jcap}\ }\textbf {\bibinfo {volume} {3}},\ \bibinfo {eid} {018}
  (\bibinfo {year} {2014})},\ \Eprint {http://arxiv.org/abs/1312.3638}
  {arXiv:1312.3638 [astro-ph.CO]} \BibitemShut {NoStop}%
\bibitem [{\citenamefont {{Percival}}\ and\ \citenamefont
  {{White}}(2009)}]{PW09}%
  \BibitemOpen
  \bibfield  {author} {\bibinfo {author} {\bibfnamefont {W.~J.}\ \bibnamefont
  {{Percival}}}\ and\ \bibinfo {author} {\bibfnamefont {M.}~\bibnamefont
  {{White}}},\ }\href {\doibase 10.1111/j.1365-2966.2008.14211.x} {\bibfield
  {journal} {\bibinfo  {journal} {\mnras}\ }\textbf {\bibinfo {volume} {393}},\
  \bibinfo {pages} {297} (\bibinfo {year} {2009})},\ \Eprint
  {http://arxiv.org/abs/0808.0003} {arXiv:0808.0003} \BibitemShut {NoStop}%
\bibitem [{\citenamefont {Buchert}\ \emph {et~al.}(1994)\citenamefont
  {Buchert}, \citenamefont {Melott},\ and\ \citenamefont {Weiss}}]{BMW94}%
  \BibitemOpen
  \bibfield  {author} {\bibinfo {author} {\bibfnamefont {T.}~\bibnamefont
  {Buchert}}, \bibinfo {author} {\bibfnamefont {A.}~\bibnamefont {Melott}}, \
  and\ \bibinfo {author} {\bibfnamefont {A.}~\bibnamefont {Weiss}},\
  }\href@noop {} {\  (\bibinfo {year} {1994})},\ \Eprint
  {http://arxiv.org/abs/astro-ph/9412075} {arXiv:astro-ph/9412075 [astro-ph]}
  \BibitemShut {NoStop}%
\bibitem [{\citenamefont {Matsubara}(2011)}]{M11}%
  \BibitemOpen
  \bibfield  {author} {\bibinfo {author} {\bibfnamefont {T.}~\bibnamefont
  {Matsubara}},\ }\href {\doibase 10.1103/PhysRevD.83.083518} {\bibfield
  {journal} {\bibinfo  {journal} {Phys. Rev. D}\ }\textbf {\bibinfo {volume}
  {83}},\ \bibinfo {pages} {083518} (\bibinfo {year} {2011})}\BibitemShut
  {NoStop}%
\bibitem [{\citenamefont {{Carlson}}\ \emph {et~al.}(2013)\citenamefont
  {{Carlson}}, \citenamefont {{Reid}},\ and\ \citenamefont {{White}}}]{CRW13}%
  \BibitemOpen
  \bibfield  {author} {\bibinfo {author} {\bibfnamefont {J.}~\bibnamefont
  {{Carlson}}}, \bibinfo {author} {\bibfnamefont {B.}~\bibnamefont {{Reid}}}, \
  and\ \bibinfo {author} {\bibfnamefont {M.}~\bibnamefont {{White}}},\ }\href
  {\doibase 10.1093/mnras/sts457} {\bibfield  {journal} {\bibinfo  {journal}
  {\mnras}\ }\textbf {\bibinfo {volume} {429}},\ \bibinfo {pages} {1674}
  (\bibinfo {year} {2013})},\ \Eprint {http://arxiv.org/abs/1209.0780}
  {arXiv:1209.0780 [astro-ph.CO]} \BibitemShut {NoStop}%
\bibitem [{\citenamefont {{Rampf}}\ and\ \citenamefont
  {{Buchert}}(2012)}]{RB12}%
  \BibitemOpen
  \bibfield  {author} {\bibinfo {author} {\bibfnamefont {C.}~\bibnamefont
  {{Rampf}}}\ and\ \bibinfo {author} {\bibfnamefont {T.}~\bibnamefont
  {{Buchert}}},\ }\href {\doibase 10.1088/1475-7516/2012/06/021} {\bibfield
  {journal} {\bibinfo  {journal} {\jcap}\ }\textbf {\bibinfo {volume} {6}},\
  \bibinfo {eid} {021} (\bibinfo {year} {2012})},\ \Eprint
  {http://arxiv.org/abs/1203.4260} {arXiv:1203.4260 [astro-ph.CO]} \BibitemShut
  {NoStop}%
\bibitem [{\citenamefont {Reid}\ and\ \citenamefont {White}(2011)}]{RW11}%
  \BibitemOpen
  \bibfield  {author} {\bibinfo {author} {\bibfnamefont {B.~A.}\ \bibnamefont
  {Reid}}\ and\ \bibinfo {author} {\bibfnamefont {M.}~\bibnamefont {White}},\
  }\href {\doibase 10.1111/j.1365-2966.2011.19379.x} {\bibfield  {journal}
  {\bibinfo  {journal} {Mon.Not.Roy.Astron.Soc.}\ }\textbf {\bibinfo {volume}
  {417}},\ \bibinfo {pages} {1913} (\bibinfo {year} {2011})},\ \Eprint
  {http://arxiv.org/abs/1105.4165} {arXiv:1105.4165 [astro-ph.CO]} \BibitemShut
  {NoStop}%
\bibitem [{\citenamefont {Suto}\ and\ \citenamefont {Sasaki}(1991)}]{S91}%
  \BibitemOpen
  \bibfield  {author} {\bibinfo {author} {\bibfnamefont {Y.}~\bibnamefont
  {Suto}}\ and\ \bibinfo {author} {\bibfnamefont {M.}~\bibnamefont {Sasaki}},\
  }\href {\doibase 10.1103/PhysRevLett.66.264} {\bibfield  {journal} {\bibinfo
  {journal} {Phys. Rev. Lett.}\ }\textbf {\bibinfo {volume} {66}},\ \bibinfo
  {pages} {264} (\bibinfo {year} {1991})}\BibitemShut {NoStop}%
\bibitem [{\citenamefont {Carlson}\ \emph {et~al.}(2009)\citenamefont
  {Carlson}, \citenamefont {White},\ and\ \citenamefont {Padmanabhan}}]{CWP09}%
  \BibitemOpen
  \bibfield  {author} {\bibinfo {author} {\bibfnamefont {J.}~\bibnamefont
  {Carlson}}, \bibinfo {author} {\bibfnamefont {M.}~\bibnamefont {White}}, \
  and\ \bibinfo {author} {\bibfnamefont {N.}~\bibnamefont {Padmanabhan}},\
  }\href {\doibase 10.1103/PhysRevD.80.043531} {\bibfield  {journal} {\bibinfo
  {journal} {Phys.Rev.}\ }\textbf {\bibinfo {volume} {D80}},\ \bibinfo {pages}
  {043531} (\bibinfo {year} {2009})},\ \Eprint {http://arxiv.org/abs/0905.0479}
  {arXiv:0905.0479 [astro-ph.CO]} \BibitemShut {NoStop}%
\end{thebibliography}%

\begin{widetext}

\appendix

\section{Power spectra}
\label{AppPower}
\subsection{Velocity divergence and vorticity}

\begin{subequations}
\begin{align} 
P_{\bar\theta\bar\theta,22}(k) &=[(2\pi)^3 \delta_{\rm D}(0) ]^{-1} \langle \bar \theta_2(\v{k})  \bar \theta_2(-\v{k})  \rangle\\
\notag &=\int \frac{\vol{3}{p_1}\vol{3}{p_2}\vol{3}{\tilde p_1} \vol{3}{\tilde p_2}}{(2\pi)^9}\,\frac{\delta_{\rm D}(\vp_1+\vp_2-\v{k}\,)\delta_{\rm D}(\tilde\vp_1+\tilde\vp_2+\v{k}\,)}{ \delta_{\rm D}(0)} \bar{G}_2^{(s)}(\vp_1,\vp_2)\bar{G}_2^{(s)}(\tilde\vp_1,\tilde\vp_2) \langle \delta_1(\vp_1)\delta_1(\vp_2) 
 \delta_1(\tilde\vp_1)\delta_1(\tilde\vp_2) \rangle \\
&= 2\int \frac{\vol{3}{p}}{(2\pi)^3} \left[\bar{G}_2^{(s)}(\vp,\vk-\vp)\right]^2 P_L(p)P_L(|\vk-\vp|)  \\
\notag &= \frac{k^3}{2\pi^2} \int_0^\infty dr\ \int_{-1}^1 dx 
 \frac{e^{-\sigx^2k^2 \left(2 r^2+1\right) }}{196 \left(r^2-2 r x+1\right)^2}  \left[2 \left(10 rx^2-3r-7 x\right) e^{\sigx^2k^2 r^2}+7 \left(-2 r x^2+r+x\right) e^{\sigx^2k^2 rx}\right]^2 \times \\
\notag&\qquad\qquad \qquad\qquad \qquad \times   P_L(kr)P_L\left(k\sqrt{1-2rx+r^2}\right)  \\
&= \frac{k^3}{2\pi^2} \int_0^\infty dr\ \int_{-1}^1 dx\ 
\frac{\left[2 \left( 10 rx^2-3r-7 x\right) e^{\sigx^2k^2 (r^2-rx)}+7 \left(-2 r x^2+r+x\right) \right]^2}{196 \left(r^2-2 r x+1\right)^2}     \bar P_L(kr)\bar P_L\left(k\sqrt{1-2rx+r^2}\right)  \\
P_{\theta\theta,22}(k) &= \frac{k^3}{2\pi^2} \cdot \frac{1}{196} \int_0^\infty dr\ \int_{-1}^1 dx\ 
\frac{\left(6rx^2+r-7 x\right)^2}{\left(r^2-2 r x+1\right)^2}  P_L(kr) P_L\left(k\sqrt{1-2rx+r^2}\right)  
\end{align}
\end{subequations}
\begin{subequations}
\begin{align} 
P_{\bar\theta\bar\theta,13}(k) &= 2\cdot [(2\pi)^3 \delta_{\rm D}(0) ]^{-1} \langle \bar \theta_1(\v{k})  \bar \theta_3(-\v{k})  \rangle\\
\notag &= 2\int \frac{\vol{3}{p_1}\vol{3}{p_2}\vol{3}{p_3} \vol{3}{\tilde p_1}}{(2\pi)^9}\,\frac{\delta_{\rm D}(\vp_1+\vp_2+\vp_3-\v{k}\,)\delta_{\rm D}(\tilde\vp_1+\v{k}\,)}{{ \delta_{\rm D}(0)}} \bar{G}_3^{(s)}(\vp_1,\vp_2,\vp_3)\bar{G}_1(\tilde\vp_1) \langle \delta_1(\vp_1)\delta_1(\vp_2) 
\delta_1(\vp_3)\delta_1(\tilde\vp_1) \rangle \\
 &= 6 P_L(k)\int \frac{\vol{3}{p}}{(2\pi)^3}\, \bar{G}_3^{(s)}(\vp,-\vp,\vk)\bar{G}_1(-\vk) P_L(p)\\
 %
 %
\notag  &= \frac{k^3}{2\pi^2} \bar P_L(k)\int_0^\infty dr \ \bar P_L(kr)\frac{e^{-\sigx^2k^2 r }}{168 r^3}  \Bigg\{ 3\left(r^2-1\right)^3 \Bigg[3 r^2 e^{\frac{1}{2}\sigx^2k^2 (r+1)^2} \left(\text{Ei}\left[-\tfrac{1}{2} \sigx^2k^2 (r-1)^2 \right]-\text{Ei}\left[-\tfrac{1}{2} \sigx^2k^2 (r+1)^2 \right]\right) \\
\notag &\qquad\qquad\qquad\qquad\qquad\qquad\qquad\qquad\qquad \quad -\left(7 r^2+2\right)  \log \left|\frac{r-1}{r+1}\right| e^{\sigx^2k^2 r (r+1)}\Bigg]  \\
\notag &\qquad\qquad\qquad\qquad\qquad\qquad\qquad + 336 r^5 e^{\sigx^2k^2 r }-2 r \left(21 r^6-50 r^4+79 r^2-6\right)e^{\sigx^2k^2 r (r+1) }\\
 &\qquad\qquad\qquad\qquad\qquad\qquad\qquad+\frac{6r^2 (r^2-1) }{(\sigx k)^2} \left[e^{2 \sigx^2k^2 r} (3 r^2 + 6r -25)-(3r^2-6r-25)\right] \\
\notag &\qquad\qquad\qquad\qquad\qquad\qquad\qquad+ \frac{12r}{(\sigx k)^4}  \left[(3r^3-12r^2-19r-28)- e^{2\sigx^2k^2 r} (3 r^3+12r^2-19r+28)\right]  \\
\notag &\qquad\qquad\qquad\qquad\qquad\qquad\qquad+\frac{48}{(\sigx k)^6} \left[e^{2\sigx^2k^2 r }(3r^2-12 r+7)-(3r^2+12r+7)\right]\\
\notag &\qquad\qquad\qquad\qquad\qquad\qquad\qquad +\frac{576}{(\sigx k)^8}\left(e^{2 \sigx^2k^2 r}-1\right)\Bigg\} \\
P_{\theta\theta,13}(k)  &= \frac{k^3}{2\pi^2} \cdot \frac{1}{168} P_L(k)\int_0^\infty dr \ P_L(kr) \Bigg\{ \frac{12}{ r^2} -82 +4r^2 -6 r^4 -\frac{3}{ r^3} \left(r^2-1\right)^3 \left(r^2+2\right) \log \left|\frac{r-1}{r+1}\right| \Bigg\}
\end{align}
$\mathrm{Ei}(x)$ denotes the exponential integral defined as $\mathrm{Ei}(x) = - \int_{-x}^\infty t^{-1} e^{-t}\ dt$.

\begin{subequations}
\begin{align}
P_{\bar{\v{w}}\bar{\v{w}},22}(k) &= [(2\pi)^3 \delta_{\rm D}(0) ]^{-1}\langle \bar{\v{w}}_2(\v{k}) \cdot  \bar{\v{w}}_2(-\v{k})  \rangle\,. \\
\notag &= \int \frac{\vol{3}{p_1}\vol{3}{p_2}\vol{3}{\tilde p_1} \vol{3}{\tilde p_2}}{(2\pi)^9}\,\frac{\delta_{\rm D}(\vp_1+\vp_2-\vk\,) \delta_{\rm D}(\tilde\vp_1+\tilde\vp_2+\vk\,)}{ \delta_{\rm D}(0)}\bar{\v{W}}_2(\vp_1,\vp_2)\cdot \bar{\v{W}}_2(\tilde\vp_1,\tilde\vp_2) \langle \delta_1(\vp_1)\delta_1(\vp_2)\delta_1(\tilde\vp_1)\delta_1(\vp_2) \rangle\\
&= 2\int \frac{\vol{3}{p}}{(2\pi)^3} \left|\bar{\v{W}}_2^{(s)}(\vp,\vk-\vp)\right|^2  P_L(p)P_L(|\vk-\vp|)  \\
\notag&= \frac{k^3}{2\pi^2} \int_0^\infty dr\ \int_{-1}^1 dx \ 
\frac{\left(1-x^2\right) (1-2 r x)^2 e^{-\sigx^2 k^2 \left(2 r^2+1\right)} \left(e^{\sigx^2k^2 r^2}-e^{\sigx^2 k^2 rx}\right)^2}{4 \left(r^2-2 r x+1\right)^2} P_L(kr)P_L\left(k\sqrt{1-2rx+r^2}\right)  \\
&= \frac{k^3}{2\pi^2} \int_0^\infty dr\ \int_{-1}^1 dx \ 
\frac{\left(1-x^2\right) (1-2 r x)^2 \left(e^{\sigx^2k^2 (r^2-rx)}-1\right)^2}{4 \left(r^2-2 r x+1\right)^2} \bar P_L(kr) \bar P_L\left(k\sqrt{1-2rx+r^2}\right) \\
P_{\v{w}\v{w},22}(k) &= 0
\end{align}
\end{subequations}

In the limit $\sigx \rightarrow 0$ we recover the standard SPT kernel as given in Eqs. (5) in \cite{S91}.
\end{subequations}

\subsection{Cross spectrum between density and velocity divergence}
\begin{subequations}
\begin{align}
P_{\bar\delta\bar\theta,22}(k)  &= [(2\pi)^3 \delta_{\rm D}(0) ]^{-1}\langle \bar \delta_2(\v{k})  \bar \theta_2(-\v{k})  \rangle\\
\notag &= -\int \frac{\vol{3}{p_1}\vol{3}{p_2}\vol{3}{\tilde p_1} \vol{3}{\tilde p_2}}{(2\pi)^9}\, \frac{\delta_{\rm D}(\vp_1+\vp_2-\v{k}\,)\delta_{\rm D}(\tilde\vp_1+\tilde\vp_2+\v{k}\,)}{\delta_{\rm D}(0)} \bar{F}_2^{(s)}(\vp_1,\vp_2)\bar{G}_2^{(s)}(\tilde\vp_1,\tilde\vp_2) \langle \delta_1(\vp_1)\delta_1(\vp_2) 
 \delta_1(\tilde\vp_1)\delta_1(\tilde\vp_2) \rangle \\
&= -2\int \frac{\vol{3}{p}}{(2\pi)^3} \bar{F}_2^{(s)}(\vp,\vk-\vp)\bar{G}_2^{(s)}(\vp,\vk-\vp)  P_L(p) P_L(|\vk-\vp|)  \\
\notag &=  -\frac{k^3}{2\pi^2} \int_0^\infty dr\ \int_{-1}^1 dx \ \frac{\left(10 rx^2-3r-7 x\right) e^{-\sigx^2k^2 \left(r^2+1\right)} \left[2 \left(10 rx^2-3r-7 x\right) e^{\sigx^2k^2 r^2}+7 \left(-2 r x^2+r+x\right) e^{\sigx^2 k^2 rx}\right]}{196 \left(r^2-2 r x+1\right)^2} \\
\notag &\qquad\qquad\qquad\qquad\qquad\times P_L(kr)P_L\left(k\sqrt{1-2rx+r^2}\right)  \\
&= -\frac{k^3}{2\pi^2} \int_0^\infty dr\ \int_{-1}^1 dx \ \frac{\left(10 rx^2-3r-7 x\right) e^{\sigx^2k^2 \left(r^2-rx\right)} \left[2 \left(10 rx^2-3r-7 x\right) e^{\sigx^2k^2 (r^2-rx)}+7 \left(-2 r x^2+r+x\right)\right]}{196 \left(r^2-2 r x+1\right)^2} \\
\notag &\qquad\qquad\qquad\qquad\qquad\times \bar P_L(kr) \bar P_L\left(k\sqrt{1-2rx+r^2}\right)\\
P_{\delta\theta,22}(k)&= -\frac{k^3}{2\pi^2} \int_0^\infty dr\ \int_{-1}^1 dx \ \frac{ \left(10 rx^2-3r-7 x\right)\left(6 rx^2+r-7x \right)}{196 \left(r^2-2 r x+1\right)^2}  P_L(kr) P_L\left(k\sqrt{1-2rx+r^2}\right)
\end{align}
\end{subequations}
\begin{subequations}
\begin{align} 
  P_{\bar\delta\bar\theta,13}(k) &= [(2\pi)^3 \delta_{\rm D}(0) ]^{-1} \left( \langle \bar \delta_1(\v{k})  \bar \theta_3(-\v{k})  \rangle + \langle \bar \delta_3(\v{k})  \bar \theta_1(-\v{k})  \rangle \right)\\
\notag &= -\int \frac{\vol{3}{p_1}\vol{3}{p_2}\vol{3}{p_3} \vol{3}{\tilde p_1}}{(2\pi)^9}\,\frac{\delta_{\rm D}(\vp_1+\vp_2+\vp_3-\v{k}\,)\delta_{\rm D}(\tilde\vp_1+\v{k}\,)}{\delta_{\rm D}(0)} \times\\
 \notag&\qquad\qquad\qquad\qquad\qquad\qquad \times \left[ \bar{G}_3^{(s)}(\vp_1,\vp_2,\vp_3)\bar{F}_1(\tilde\vp_1) + \bar{F}_3^{(s)}(\vp_1,\vp_2,\vp_3)\bar{G}_1(\tilde\vp_1) \right] \langle \delta_1(\vp_1)\delta_1(\vp_2) \delta_1(\vp_3)\delta_1(\tilde\vp_1) \rangle \\
  &=-3 P_L(k)\int \frac{\vol{3}{p}}{(2\pi)^3}\, \left[ \bar{F}_1(\vk)\bar{G}_3^{(s)}(\vp,-\vp,-\vk) + \bar{F}_3^{(s)}(\vp,-\vp,\vk)\bar{G}_1(-\vk) \right]P_L(p)\\
  %
%
\notag&= -\frac{k^3}{4\pi^2} \bar P_L(k)\int_0^\infty dr\ \bar P_L(kr)
\frac{e^{-\sigx^2k^2r }}{504 r^3} \Bigg\{ 3  \left(r^2-1\right)^3 \Bigg[9 r^2 e^{\frac{1}{2} \sigx^2k^2 (r+1)^2} \left(\text{Ei}\left[-\tfrac{1}{2} \sigx^2k^2 (r-1)^2 \right]-\text{Ei}\left[-\tfrac{1}{2} \sigx^2k^2 (r+1)^2 \right]\right)\\
\notag&\qquad\qquad\qquad\qquad\qquad\qquad\qquad\qquad\qquad\quad -2 \left(7 r^2+2\right) 2\log \left|\frac{r-1}{r+1}\right| e^{\sigx^2k^2 r (r+1) }\Bigg]\\
\notag &\qquad\qquad\qquad\qquad\qquad\qquad\qquad\quad + 108 r^5  e^{K^2 r R^2}-8  r \left(21 r^6-50 r^4+79 r^2-6\right)  e^{\sigx^2k^2 r (r+1) }\\
&\qquad\qquad\qquad\qquad\qquad\qquad\qquad\quad + \frac{18r^2 (r^2-1)}{(\sigx k)^2}  \left[e^{2 \sigx^2k^2 r} (3r^2+6r-25)- (3r^2-6r-25)\right] \\
\notag&\qquad\qquad\qquad\qquad\qquad\qquad\qquad\quad+ \frac{36r}{(\sigx k)^4} \left[(3r^3-12r^2-19r-28) - e^{2 \sigx^2k^2 r}(3 r^3+12r^2-19r+28)\right] \\
\notag &\qquad\qquad\qquad\qquad\qquad\qquad\qquad\quad+ \frac{144}{(\sigx k)^6} \left[ e^{2 \sigx^2k^2 r}(3r^2-12r +7)-(3r^2+12r+7)\right]\\
\notag &\qquad\qquad\qquad\qquad\qquad\qquad\qquad\quad+ \frac{1728}{(\sigx k)^8} \left(e^{2 \sigx^2k^2 r} -1\right)\Bigg\}   \\
P_{\delta\theta,13}(k) &= -\frac{k^3}{2\pi^2} \cdot \frac{1}{504} P_L(k)\int_0^\infty dr\ P_L(kr) \Bigg\{ \frac{24}{r^2} -202 +56 r^2 -30 r^4 - \frac{3}{r^3} \left(r^2-1\right)^3 \left(5 r^2+4\right) \log \left|\frac{r-1}{r+1}\right| \Bigg\}
\end{align}
\end{subequations}
 In the limit $\sigx \rightarrow 0$ the kernels reduce to the standard SPT result given in Eqs. (A25)-(A26) in \cite{CWP09}, note however that our convention for $\theta_n$ is different compared to \cite{CWP09} such that the power spectra have flipped signs.

\section{Eulerian kernels in terms of Lagrangian kernels}
\label{AppLag}
Inserting the expression \eqref{JW} for $J_{\bar F} \bar w_i$ and the perturbative ansatz for $\v{\bar \varPsi}$, see \eqref{pertExpPsiCG}, in \eqref{wFromPsiCG}  we obtain
\begin{subequations}\label{WSTrelation}
 \begin{align} 
 -\bar{\v{W}}^{(2)}_{\rm s}(\v{p}_1,\v{p}_2) &=  2\, \v{k}\times\v{\bar T}^{(2)}(\v{p}_1,\v{p}_2) \,,\\
 -\bar{\v{W}}^{(3)}_{\rm s} (\v{p}_1,\v{p}_2,\v{p}_3)&=  3\, \v{k}\times\v{\bar T}^{(3)} + \frac{1}{3}\stackrel{\ + \text{ cyclic permutations of } (\vp_1,\vp_2,\vp_3)\  }{\Big[\Big( \v{k}_1\times \v{k}_2\ \v{\bar S}^{(1)}\cdot \Big[\v{\bar S}^{(2)}+ \v{\bar T}^{(2)}\Big]+2\,\v{k}_2\times \v{\bar T}^{(2)}\, \v{k}\cdot \v{\bar S}^{(1)}-2\,\v{\bar S}^{(1)}\,\v{k}_1\cdot \left(\v{k}_2 \times \v{\bar T}^{(2)}\right)\Big)\Big] }\,. \end{align}
\end{subequations}
 For the sake of brevity we suppress the functional dependencies on the right hand side. They can be easily restored by attaching each kernel a dependence on $(\vp_{i}, \ldots, \vp_{i+n-1})$ in ascending order beginning with $i=1$ from the left, for example $\v{\bar S}^{(1)}\cdot\v{\bar T}^{(2)}:=\v{\bar S}^{(1)}(\vp_1)\cdot\v{\bar T}^{(2)}(\vp_2,\vp_3)$. We defined $ \bar{\v{W}}^{(n)}_{\rm s}(\v{p}_1,..,\v{p}_n) := 1/n! \sum_{\sigma\in S_n} \bar{\v{W}}^{(n)}(\v{p}_{\sigma(1)},..,\v{p}_{\sigma(n)} )\,,$
 where the sum goes over all $n!$ permutations of $n$ indices. 
Analogous to the case of dust where $F_n$ is related to $\v{S}^{(n)}$ and $\v{T}^{(n)}$ via Eqs.\,(6.9) and (B1)-(B3) in \cite{RB12}, we obtain from Eqs.\,\eqref{deltathetaFromPsiCG} and \eqref{wFromPsiCG}
 \begin{subequations}
 \label{FSTrelation}
\begin{align}
\bar F_1^{\rm s}(\v{p}_1) &= \v{k} \cdot \v{\bar S}^{(1)} \,,\\
\bar F_2^{\rm s} (\v{p}_1,\v{p}_2) &= \v{k} \cdot \v{\bar S}^{(2)} +\frac{1}{2} \left(\v{k} \cdot \v{\bar S}^{(1)}\right)\left( \v{k} \cdot \v{\bar S}^{(1)} \right) \,,\\
\bar F_3^{\rm s}(\v{p}_1,\v{p}_2,\v{p}_3)  &= \v{k}\cdot \v{\bar S}^{(3)}+ \frac{1}{6} \left( \v{k} \cdot \v{\bar S}^{(1)} \right)\left(  \v{k} \cdot \v{\bar S}^{(1)} \right)\left(  \v{k} \cdot \v{\bar S}^{(1)}\right)+ \frac{1}{3}\stackrel{+ \text{cyclic permutation of } (\vp_1,\vp_2,\vp_3)}{\left(\v{k}\cdot \v{\bar S}^{(1)} \right)\left( \v{k}\cdot \left[ \v{\bar S}^{(2)}+\v{\bar T}^{(2)}\right]\right) } \,.
\end{align}
\end{subequations}

\end{widetext}

\end{document}